\documentclass[a4paper,11pt]{article}
\pdfoutput=1
\usepackage{jheppub} 
\usepackage{graphicx,hyperref,braket,orcidlink, algorithm2e, amssymb, amsmath, dsfont, subfigure, lineno, bm, slashed, multirow, sidecap}
\usepackage[nolist,nohyperlinks]{acronym}
\graphicspath{{figs/}}
\hypersetup{
	bookmarks=true,
	unicode=true,
	pdftoolbar=true,
	pdfmenubar=true,
	pdffitwindow=false,
	pdfstartview={FitH},
	pdftitle={pdf},
	pdfauthor={J.Y.~Araz, M.~Spannowsky}, 
	pdfsubject={pdf},
	pdfcreator={J.Y.~Araz, M.~Spannowsky},
	pdfproducer={}, 
	pdfkeywords={},
	pdfnewwindow=true,
	colorlinks=true,
	linkcolor=blue,
	citecolor=olive,
	filecolor=magenta,
	urlcolor=cyan
}

\newcommand{\Qtyl}{\hat{q}_{1-\alpha}}

\title{Another Fit Bites the Dust:\\
Conformal Prediction as a Calibration Standard for Machine Learning in High-Energy Physics}

\author[1,2]{Jack Y. Araz\orcidlink{0000-0001-8721-8042}}
\emailAdd{j.araz@ucl.ac.uk}
 
\affiliation[1]{Department of Physics and Astronomy, University College London, London, WC1E 6B, UK}
\affiliation[2]{Department of Engineering, City St. George's, University of London, London, EC1V 0HB, UK}

\author[3]{Michael Spannowsky\orcidlink{0000-0002-8362-0576}}
\emailAdd{michael.spannowsky@durham.ac.uk}
\affiliation[3]{Institute for Particle Physics Phenomenology and Department of Physics, Durham University, Durham DH1 3LE, UK}

\preprint{IPPP/25/88} 

\abstract{Machine-learning techniques are essential in modern collider research, yet their probabilistic outputs often lack calibrated uncertainty estimates and finite-sample guarantees, limiting their direct use in statistical inference and decision-making. Conformal prediction (CP) provides a simple, distribution-free framework for calibrating arbitrary predictive models without retraining, yielding rigorous uncertainty quantification with finite-sample coverage guarantees under minimal exchangeability assumptions, without reliance on asymptotics, limit theorems, or Gaussian approximations. In this work, we investigate CP as a unifying calibration layer for machine-learning applications in high-energy physics. Using publicly available collider datasets and a diverse set of models, we show that a single conformal formalism can be applied across regression, binary and multi-class classification, anomaly detection, and generative modelling, converting raw model outputs into statistically valid prediction sets, typicality regions, and $p$-values with controlled false-positive rates. 
While conformal prediction does not improve raw model performance, it enforces honest uncertainty quantification and transparent error control. We argue that conformal calibration should be adopted as a standard component of machine-learning pipelines in collider physics, enabling reliable interpretation, robust comparisons, and principled statistical decisions in experimental and phenomenological analyses.}

\begin{document}

\maketitle

\section{Introduction}

The search for new phenomena at the Large Hadron Collider (LHC) relies on extracting weak signals from overwhelming backgrounds, which are subject to significant statistical and systematic uncertainties. Machine learning (ML) has become central to this effort, enabling the construction of powerful discriminants for classification~\cite{Chen:2023ind, Nachman:2021yvi, Baldi:2016fzo, Jang:2025guj, Bassa:2025qru, Kheddar:2024osf, Han:2023djl, Choi:2023slq, Araz:2021zwu, Bols:2020bkb, Ferriere:2025csu}, 
regression~\cite{Piscopo:2019txs,Araz:2021hpx,Algren:2024bqw, Kim:2023koz, CRESST:2022qor, Maier:2021ymx, Carrazza:2019efs, Buckley:2025pqk}, and anomaly detection tasks \cite{Kasieczka:2021xcg, Nachman:2020lpy, Araz:2025oax, Ngairangbam:2025fst,Atkinson:2022uzb,DAgnolo:2018cun, Collins:2018epr, Collins:2019jip, DAgnolo:2019vbw, Farina:2018fyg, Heimel:2018mkt, Blance:2019ibf, Hajer:2018kqm, Baldi:2014pta, Komiske:2019aa}. Despite these advances, a central challenge remains: while modern ML models achieve remarkable accuracy, their predictions are often poorly calibrated, providing no guarantees of uncertainty coverage or false-positive control. As a result, it is difficult to quantify the reliability of model outputs, particularly when training data and test data differ or when signals populate regions unseen during training. In an experimental setting, this uncertainty in prediction reliability directly translates into ambiguity in interpreting the results. This limitation has been opened up the research into uncertainty estimation~\cite{Romero:2025rck, Keller:2025bac, Peron:2025mtj, Benevedes:2025nzr, Desai:2025mpy, Elsharkawy:2025yeb, Khot:2025kqg, Kriesten:2024ist, Panahi:2024sfb, Bieringer:2024nbc, Dickinson:2023yes, Golutvin:2023fle, Koh:2023wst, Cheung:2022dil, Bellagente:2021yyh, Barnard:2016qma, Nachman:2019yfl, Nachman:2019dol} and mitigation~\cite{SuperCDMS:2025ywe, Azakli:2025yfb, Stein:2022nvf, Araz:2021wqm, Louppe:2016ylz, Englert:2018cfo} in the setting of high-energy physics for machine learning applications.

Traditional approaches in high-energy physics classification tasks often interpret the output of a neural network or boosted decision tree as a calibrated probability and apply fixed thresholds to classifier scores to define signal- or background-like regions. However, such thresholds implicitly assume that the model's predicted probabilities are well-calibrated and that the training and test distributions coincide, assumptions that rarely hold in practice. Techniques such as isotonic regression~\cite{doi:10.1287/moor.14.2.303} or Platt scaling~\cite{Platt1999} can improve calibration, but do not guarantee coverage when the model is applied to shifted or mixed data domains. 

Conformal prediction\footnote{We use the term conformal prediction to denote the construction of prediction sets and intervals with guaranteed coverage, while conformal inference refers more broadly to the use of conformal methods for statistically valid uncertainty quantification and hypothesis testing.} provides a statistically principled approach to address this challenge. It is a general, model-agnostic framework that transforms arbitrary predictive models into calibrated systems with finite-sample guarantees~\cite{Vovk:2005, Shafer:2008, Lei:2018, 10.1007/3-540-36755-1_29, https://doi.org/10.1111/rssb.12021, angelopoulos2022gentleintroductionconformalprediction}. 
CP has already been used for problems in language domain~\cite{kaur2025empiricalstudyconformalprediction, fisch2021efficientconformalpredictioncascaded, fisch2021fewshotconformalpredictionauxiliary}, token level predictions~\cite{dey2021conformalpredictiontextinfilling}, image classification~\cite{Angelopoulos:2021, kutiel2023conformal}, time-serries forecasting~\cite{xu2020enbpi}, anomaly detection~\cite{Hennh_fer_2024, xu2021ecad}, out-of-order distribution testing~\cite{liang2024integrative}, in clinical medicine~\cite{Vazquez:2022aa}, medical imaging~\cite{lu2022fairconformalpredictorsapplications}, biological sequence annotation~\cite{10.1371/journal.pcbi.1012135}, large-scale protein search~\cite{Boger:2025aa}, and computer vision \cite{Romano:2019, Angelopoulos:2021}, it remains largely unexplored in high-energy physics. Yet its defining characteristics align closely with the methodological needs of collider analyses. First, CP provides coverage guarantees that hold for finite datasets, thereby circumventing asymptotic assumptions often used in likelihood-based uncertainty estimates. Second, it directly calibrates residuals or anomaly scores, yielding interpretable p-values or intervals without relying on parametric noise models. Third, CP enables local or conditional calibration: through techniques such as Conformalised Quantile Regression (CQR)~\cite{Romano:2019}, Adaptive CP~\cite{Papadopoulos:2008, Barber:2021}, and Mondrian CP~\cite{Vovk:2003}, one can maintain coverage within specific regions of phase space, for example, as a function of invariant mass or transverse momentum. This property naturally complements the differential validation strategies used in collider analyses.

Methodologically, this study aims to bridge established HEP validation practices with recent advances in predictive inference. In practice, it illustrates, through various examples, that CP can improve the interpretability and statistical robustness of ML-driven analyses without access to proprietary experimental calibration frameworks. We show that CP can act as a unifying statistical layer for the reliable deployment of ML methods in particle physics, complementing rather than replacing traditional frequentist and Bayesian methods. By integrating conformal calibration into existing ML workflows in collider physics, we demonstrate that rigorous uncertainty quantification is achievable across diverse analysis settings.

\begin{figure}[h]
    \centering
    \includegraphics[width=\linewidth]{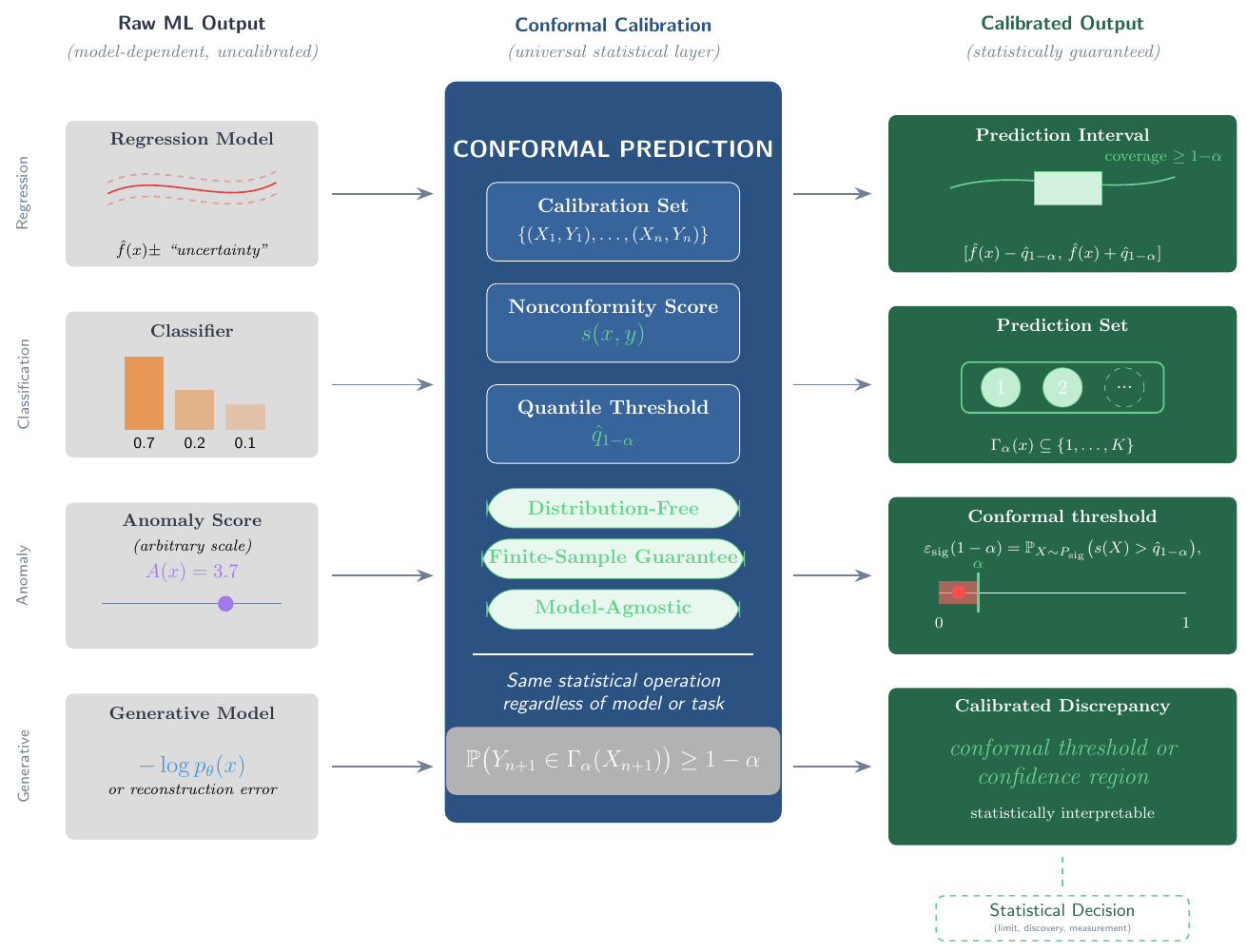}
    \caption{Schematic overview of conformal prediction as a universal calibration layer for HEPML. A model-specific, uncalibrated output is combined with a calibration sample $\mathcal{D}_{\rm cal}=\{(X_i,Y_i)\}_{i=1}^{n_{\rm cal}}$ through a chosen nonconformity score $s(x,y)$ to determine the split-conformal threshold $\hat{q}_{1-\alpha}$. This results in calibrated objects with finite-sample marginal guarantees: prediction intervals in regression, label sets $\Gamma_\alpha(x)\subseteq\{1,\dots,K\}$ in classification, and calibrated anomaly or generative-model discrepancies that can be expressed as conformal $p$-values or, equivalently, as threshold scores.}
    \label{fig:summary-fig}
\end{figure}

The paper's central claim is that conformal prediction provides a missing statistical layer for machine learning in high-energy physics: it converts arbitrary ML outputs into calibrated, distribution-free statements with finite-sample guarantees, without modifying the underlying model or training procedure; see Fig.~\ref{fig:summary-fig} for a schematic overview. We argue that conformal calibration should become a standard post-processing step for ML-based collider analyses. Beyond the specific case studies presented here, we view this work as a call to the high-energy physics community to adopt CP as a standard component of machine-learning workflows. As ML models become increasingly central to experimental analyses, the absence of distribution-free uncertainty guarantees poses a growing risk for robustness and interpretability. 

These properties make CP particularly attractive for HEPML, where high-dimensional observables, imperfect simulations, and subtle distribution shifts can compromise the interpretability of raw model outputs. Our goal is not to propose new architectures, but to demonstrate, across representative tasks in collider phenomenology, that a simple conformal calibration layer turns arbitrary ML outputs into objects with a clear frequentist meaning. This suggests a concrete community standard: when reporting headline performance metrics such as AUC, background rejection, or mean-squared error, analyses should also report conformal reliability diagnostics, including empirical coverage versus nominal $1-\alpha$, efficiency measures such as mean interval width or mean prediction-set size, and conditional coverage versus key kinematic variables. Adopting these checks would make model comparisons more meaningful and help translate ML performance into robust analysis decisions under controlled error rates.

This paper is organised as follows: Section~\ref{sec:theory} introduces the core idea of conformal prediction. Section~\ref{sec:regression} starts from a regression problem to demonstrate conformal inference on a toy example. Section~\ref{sec:datasets} then summarises the public collider datasets used in this study. In Section~\ref{sec:classification}, we apply conformal prediction to both binary and multi-class jet tagging. Section~\ref{sec:anomaly} turns to unsupervised anomaly detection and shows how conformal calibration maps arbitrary anomaly scores to calibrated $p$-values with controlled background false-positive rate. Section~\ref{sec:generative} extends CP to generative settings to obtain calibrated discrepancy measures from model outputs. We conclude in Section~\ref{sec:conclusion} with a discussion of practical caveats, guidance on interpretation, and an outlook.

\section{Conformal prediction for predictive inference}
\label{sec:theory}

Conformal prediction (CP) is a general framework for constructing prediction sets with rigorous, distribution-free finite-sample guarantees~\cite{Vovk:2005, Shafer:2008}. The central and unifying principle of CP is the guarantee of {\it marginal coverage}, which holds for any predictive task, including regression, classification, density-based modelling, or anomaly detection, provided the observed data are exchangeable.

Let
\(
\{Z_i\}_{i=1}^n = \{(X_i, Y_i)\}_{i=1}^n
\)
be a sequence of random variables taking values in a measurable space $\mathcal{Z}=\mathcal{X}\times\mathcal{Y}$. We assume that the data are exchangeable, which holds in particular when the samples are independent and identically distributed (i.i.d.). Exchangeability implies that the joint distribution of the data is invariant under permutations of the indices and constitutes the sole probabilistic assumption required by CP.
Given any measurable nonconformity score $s:\mathcal{X}\times\mathcal{Y}\to\mathbb{R}$, one can quantify how atypical an observation $Z=(x,y)$ is relative to the rest of the data. Importantly, CP imposes no restrictions on the functional form of $s$. It may depend on a trained machine-learning model, a likelihood function, or a domain-specific statistic. The flexibility in choosing $s$ enables CP to act as a wrapper around arbitrary predictive or generative models.

In its simplest form, conformal prediction constructs a prediction set for a future observation $Y_{n+1}$ at a given input $X_{n+1}$ by comparing the nonconformity score of candidate outputs to the empirical distribution of scores computed from the observed data. In the split conformal setting, the data are divided into a training set, used to construct any auxiliary objects such as a predictor $\hat{f}$, and a calibration set
\(
\mathcal{D}_{\mathrm{cal}} = \{Z_i\}_{i=1}^{n_{\mathrm{cal}}}
\).
Fixing the score function $s$, one computes calibration scores
\(
s_i = s(X_i, Y_i)
\)
and defines the empirical quantile
\begin{equation}
    \Qtyl
    =
    \text{the } \left\lceil (1-\alpha)(n_{\mathrm{cal}}+1) \right\rceil\text{-th smallest value of } \{s_i\}\label{eq:score-quantile}
\end{equation}
with $\alpha\in[0,1]$. Because the calibration scores are discrete and finite in number, it is essential to use the conservative quantile (ceiling) or the \emph{p}-value formulation to ensure the finite-sample inequality. Here, $1-\alpha$ is the nominal coverage level.

The conformal prediction set is then defined abstractly as
\begin{equation}
    \Gamma_\alpha(X_{n+1})
    =
    \left\{
    y \in \mathcal{Y} :
    s(X_{n+1}, y) \le \Qtyl
    \right\}.\label{eq:predset}
\end{equation}
Crucially, the validity of this construction does not depend on the interpretation of $y$ or on the particular learning task. Under the sole assumption of exchangeability, the resulting prediction set satisfies the marginal coverage guarantee
\begin{equation}
    \mathbb{P}\bigl( Y_{n+1} \in \Gamma_\alpha(X_{n+1}) \bigr) \ge 1 - \alpha,\label{eq:marginal-cov}
\end{equation}
where the probability is taken over both the calibration data and the new test point.

This guarantee has a direct interpretation familiar from frequentist uncertainty quantification in high-energy physics. 
If the same analysis were repeated over many independent pseudo-experiments drawn from the same underlying process, the conformal prediction set would contain the true outcome in at least a fraction $1-\alpha$ of those repetitions. 
Importantly, this statement is made \emph{before} observing the test data and does not rely on the correctness of the predictive model, the dimensionality of the feature space, or the form of the data-generating distribution.

This marginal coverage property constitutes the theoretical foundation of conformal prediction. It ensures that, averaged over the joint distribution of inputs and outputs, the prediction sets achieve the desired coverage level irrespective of model misspecification, dimensionality, or the complexity of the underlying data distribution. Specific applications, such as regression or classification, arise only from the choice of a nonconformity score and do not affect the validity of the coverage guarantee.

It is essential to stress what conformal prediction does and does not provide. Conformal prediction does not improve class separability, sharpen regression accuracy, or improve the performance of a poorly trained model. 
Instead, it enforces honest uncertainty quantification. In particular, any information present in the underlying model is converted into prediction sets or calibrated statistics, whose coverage properties are guaranteed by construction. If the base model is weak, the resulting conformal prediction sets will be correspondingly large or uninformative, reflecting genuine epistemic uncertainty.

Since its initial formulation, conformal prediction has seen significant theoretical development aimed at expanding its applicability, improving efficiency, and refining its coverage guarantees. While the foundational result guarantees marginal coverage under exchangeability, subsequent research has explored extensions tailored to specific learning tasks, data regimes, and structural assumptions.

In classification problems, conformal prediction naturally yields set-valued predictors that may include multiple labels when the data are ambiguous. Early studies showed that conformal classifiers achieve finite-sample marginal coverage without assumptions about the classifier or class-conditional distributions \cite{Vovk:2005, Shafer:2008}. A common approach uses scores based on class probabilities or margins produced by a base classifier, yielding prediction sets of the form $\Gamma_\alpha(X) \subseteq \mathcal{Y}$ that adjust in size according to local uncertainty.
Subsequent theoretical progress concentrated on efficiency and adaptivity. Ref.~\cite{Romano2020} introduced conformalised multiclass classifiers that control coverage while minimising expected set size, demonstrating optimality under mild regularity conditions. Related research linked conformal classification to multiple-hypothesis testing and false-discovery-rate control, elucidating the statistical interpretation of label sets~\cite{Lei:2013}.

In regression, the classical split conformal method yields prediction intervals with guaranteed marginal coverage but does not ensure conditional coverage given the input $X$. A series of impossibility results indicated that distribution-free conditional coverage is unattainable in general without trivialising the prediction sets~\cite{bibLei}. This spurred extensive work on {\it approximate} or {\it localised} conditional guarantees.

Notable methods include locally weighted conformal prediction \cite{Tibshirani:2019}, which adjusts interval width according to covariate-dependent noise levels, and quantile-based conformal methods \cite{Romano:2019} that utilise conditional quantile regression to attain asymptotically sharp prediction intervals under mild smoothness assumptions. Other strategies explore conformal prediction under weaker notions of invariance or partial exchangeability, providing robustness guarantees in time-series or grouped-data settings~\cite{angelopoulos2022gentleintroductionconformalprediction}. While such guarantees are inherently weaker than those in the classical i.i.d. case, they greatly broaden the practical scope of CP.

In the context of high-energy physics, this distribution-free guarantee is particularly attractive, as it enables principled uncertainty quantification even in the presence of imperfect simulations, complex detector effects, or high-dimensional feature spaces, provided the exchangeability assumption is approximately satisfied. 

In the following sections, we make these abstract principles concrete by applying conformal prediction to regression, classification, anomaly-detection and generative tasks relevant to collider physics.

\section{Regression with heteroscedastic uncertainty}
\label{sec:regression}

\begin{figure}[ht]
    \begin{minipage}{0.55\textwidth}
    \includegraphics[width=\linewidth]{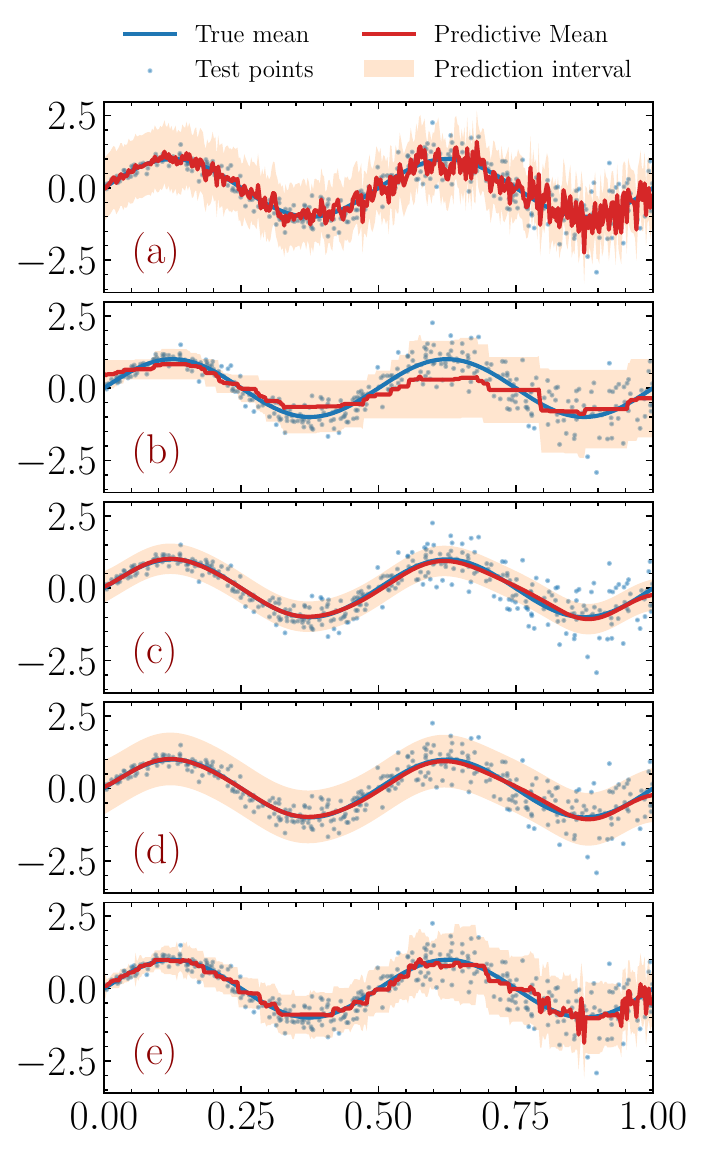}
    \end{minipage}
    \hfill
    \begin{minipage}{0.45\textwidth}
    \caption{Predictive intervals for a heteroscedastic synthetic regression task. 
    The blue curve shows the ground-truth $f(x)=\sin(4\pi x)$ and blue points are test samples generated with input-dependent noise $\sigma(x)=0.1+0.6x$. Solid red curves denote the predictive mean, and the shaded orange regions represent prediction intervals. \textbf{Panel (a)} shows split conformal prediction applied to a Random Forest regressor. \textbf{Panel (b)} shows conformalised Quantile Regression (CQR) using two 
    Gradient-Boosting quantile models. \textbf{Panel (c)} shows the effect of Gaussian Process (GP) intervals. 
    \textbf{Panel (d)} shows conformal calibration of the GP using standardised residual scores. 
    \textbf{Panel (e)} shows adaptive Conformal Prediction (ACP).\label{fig:regression}}
    \end{minipage}
\end{figure}

As a demonstration, we begin our study with regression tasks that exhibit input-dependent (heteroscedastic) noise. Using synthetic data with known ground truth, we compare naïve split CP, conformalised quantile regression, adaptive conformal prediction, and Gaussian process regression with and without conformal calibration. We use the benchmark $y=\sin(4\pi x)+\epsilon$ with $\epsilon\sim\mathcal N(0,\sigma^2(x))$ and $\sigma(x)=0.1+0.6x$. We evaluate empirical coverage, mean interval width, and interval scores across a range of miscoverage levels $\alpha$. 

Figure~\ref{fig:regression} compares a range of regression models and 
uncertainty quantification strategies, illustrating how different approaches 
behave under heteroscedastic noise. In each panel, the blue curve denotes the 
true conditional mean, the red curve is the learned predictive mean, the blue 
dots indicate the test samples, and the orange band shows the estimated 
prediction interval (PI). The quality of these intervals can be assessed 
visually by whether they expand and contract in regions where the noise level 
varies.

Figure~\ref{fig:regression}(a) shows the split conformal interval constructed 
on top of a Random Forest regressor~\cite{Breiman:2001aa, Geurts:2006aa}. 
Because this method uses absolute residuals on the calibration set and applies a single global quantile correction, the resulting interval has nearly constant width across the input domain. This highlights a key limitation of residual-based conformalisation on models that do not explicitly model heteroscedasticity: the interval becomes too wide in regions with low noise (low-$x$) and too narrow where the noise increases, thereby overestimating uncertainty in the former and 
underestimating in the latter, despite achieving marginal coverage.

In Fig.~\ref{fig:regression}(b) we present Conformalised Quantile Regression (CQR) using Gradient Boosting~\cite{bibFriedman}. CQR trains two quantile regressors to estimate lower and upper conditional quantiles before applying a small calibration correction. This produces locally adaptive intervals that contract in low-variance regions and expand in regions of higher variability. As a result, CQR avoids the overestimation observed in panel~(a) and provides a much closer match to the true heteroscedastic structure 
of the data.

Panel~(c) shows the nominal Gaussian Process (GP) interval~\cite{bibRasmussen}, 
using the model's predictive variance directly. While the GP mean captures the general trend, its modelling assumptions yield intervals that are tight in the low-$x$ region, where the kernel fits the data well, but too narrow in the high-$x$ region, where the GP underestimates the noise amplitude. This illustrates the classical issue that GP uncertainty is only well-calibrated when its assumptions about smoothness and noise variance match the data-generating process.

To address this miscalibration, Fig.~\ref{fig:regression}(d) applies conformal calibration on top of the GP using the standardised residual score
\[
  s(x,y) = \frac{|y - \mu_{\rm GP}(x)|}{\sigma_{\rm GP}(x)} \, .
\]
This rescales the GP intervals to enforce finite-sample coverage without altering the predictive mean. The resulting intervals retain local adaptivity via $\sigma_{\rm GP}(x)$ while being sufficiently broadened in regions where the nominal GP underestimates uncertainty.

Finally, Fig.~\ref{fig:regression}(e) displays Adaptive Conformal Prediction (ACP)~\cite{amoukou2023adaptiveconformalpredictionreweighting}, which learns a conditional mean predictor $\mu(x)$ together with a scale function $\hat\sigma(x)$ that models the local residual magnitude (e.g.\ the conditional median absolute deviation). Conformal calibration is performed on the standardised residuals $r_i = |y_i - \mu(x_i)| / \hat\sigma(x_i)$, and the resulting prediction interval takes the form
\[
\bigl[\mu(x) - \hat q_{1-\alpha}\,\hat\sigma(x),\; \mu(x) + \hat q_{1-\alpha}\,\hat\sigma(x)\bigr]
\]
where $q$ is the calibrated quantile. This method explicitly models 
heteroscedasticity and uses conformalisation to correct any remaining 
miscalibration. As shown in the figure, ACP produces smooth, data-adaptive intervals that closely track the true noise structure.

Overall, the figure highlights a central message of conformal prediction for regression: while classical models such as Random Forests and Gaussian Processes require strong modelling assumptions to produce well-calibrated uncertainties, conformal calibration can correct systematic miscalibration and, when combined with heteroscedastic models, yields adaptive, distribution-free prediction intervals.

\section{HEP datasets}
\label{sec:datasets}

In the rest of this study, we used publicly available collider datasets. This section summarises the details about these datasets.

The \textbf{Top Quark Tagging Dataset} is a publicly released Monte-Carlo simulated sample designed for developing and benchmarking machine-learning algorithms for boosted top tagging, as described in Ref.~\cite{10.21468/SciPostPhys.7.1.014} and made available through the Zenodo record~\cite{kasieczka_2019_2603256}. The dataset contains approximately 1.2 million training events and 400,000 events each for validation and testing, with each jet labelled as either a hadronically decaying top quark or a QCD background jet. Events were generated at a centre-of-mass energy of 14~TeV using \textsc{Pythia\,8}~\cite{Bierlich:2022pfr} for hard scattering and parton showering, followed by a fast detector simulation using \textsc{Delphes}~\cite{michele_selvaggi_2023_10034640} with an ATLAS detector configuration, and without modelling pile-up or multi-parton interactions. Jets are clustered with the anti--$k_{T}$~\cite{Cacciari:2008gp} algorithm with radius parameter $R=0.8$ in the transverse-momentum range $550 < p_{T} < 650$~GeV and pseudorapidity $|\eta|<2$. For each jet, the dataset provides the four-momentum vectors of up to the leading 200 jet constituents (zero-padded when fewer constituents are present). 

The \textbf{JetClass} dataset is a large-scale simulated dataset, released through the Zenodo record~\cite{qu_2022_6619768} and introduced in Ref.~\cite{Qu:2022mxj}. It comprises 100 million jets for training, 5 million for validation, and 20 million for testing, spanning 10 distinct jet classes representing different underlying particle types and production processes. The events were simulated using a chain of physics tools, including MadGraph~\cite{Alwall:2014hca} for hard-process generation, \textsc{Pythia\,8} for parton showering and hadronisation, and \textsc{Delphes} for fast detector simulation. 

The \textbf{ATLAS OmniFold 24-Dimensional $Z+$jets Open Data} release is a public dataset containing the unbinned $pp\to Z(\to \mu\bar{\mu})+{\rm jets}$ events, multidimensional differential cross‑section measurement of twenty‑four kinematic observables in $Z+$jets events collected with the ATLAS detector at $\sqrt{s}=13$~TeV, as presented in Ref.~\cite{PhysRevLett.133.261803} and archived on Zenodo~\cite{atlas_collaboration_2024_11507450}. This dataset represents the outcome of an OmniFold analysis that simultaneously unfolds detector effects to obtain particle‑level distributions for $Z\to\mu\mu$ events in association with jets from 139~fb$^{-1}$ of LHC proton–proton collision data. The open data contain the 24 measured particle‑level observables, including the transverse momentum, rapidity, and azimuth of the two muons, the dimuon system kinematics, and the four‑momentum and substructure observables of the two leading charged particle jets, along with accompanying Monte Carlo predictions for both "pseudo‑data" validation samples and real data. The process is defined in a boosted fiducial region with $p_{T}^{\mu\mu}>200$~GeV.  

\section{Conformal prediction for classification}
\label{sec:classification}

At first glance, one might wonder how on earth one would set guaranteed uncertainty bands in a classification setting. Yet, CP can be extended to classification by producing \emph{set-valued} predictions (label sets) rather than scalar uncertainty intervals. Let $\mathcal{Y}=\{1,\dots,K\}$ denote a finite label space of size $K$. Given a trained probabilistic classifier $\hat f:\mathcal{X}\to[0,1]^K$ satisfying $\sum_{y=1}^K \hat f_y(x)=1$, we write $\hat f_y(x)$ for the predicted probability assigned to class $y$ at input $x$. Conformal prediction converts these scores into a calibrated prediction set $\Gamma_\alpha(X_{n+1})\subseteq\mathcal{Y}$ which satisfies the finite-sample \emph{marginal coverage} guarantee in Eq.~\eqref{eq:marginal-cov}, under the sole assumption that the pooled calibration and test data are exchangeable.

The construction begins with the choice of a nonconformity score $s:\mathcal{X}\times\mathcal{Y}\to\mathbb{R}$, which measures how atypical a candidate label $y$ is at input $x$ relative to the observed data. The score may depend on a trained classifier $\hat f$, which is held fixed during calibration. A simple and widely used choice in classification is
\begin{equation}\label{eq:score_prob}
s(x,y) \;=\; 1 - \hat{f}_y(x),
\end{equation}
so that large scores correspond to labels with low predicted probability. Importantly, conformal validity does not require $\hat f$ to be well-specified as a probabilistic model; miscalibration affects the efficiency (size) of the resulting prediction sets but not the marginal coverage guarantee.

To implement split conformal prediction, the available data are partitioned into a training set, used to fit $\hat f$, and a calibration set $\mathcal{D}_{\mathrm{cal}}=\{(X_i,Y_i)\}_{i=1}^{n_{\mathrm{cal}}}$. Fixing the score function $s$, one computes calibration scores $s_i = s(X_i,Y_i), \ i=1,\dots,n_{\mathrm{cal}}$. Let $\hat q_{1-\alpha}$ denote the empirical $(1-\alpha)$-quantile of $\{s_i\}$, defined in the conservative finite-sample manner as in Eq.~\eqref{eq:score-quantile}. The conformal prediction set for a new input $X_{n+1}$ is then given by
\begin{equation}\label{eq:predset_class}
\Gamma_\alpha(X_{n+1})
\;=\;
\{\, y\in\mathcal{Y} : s(X_{n+1},y)\le \hat q_{1-\alpha} \,\}.
\end{equation}
For the score in Eq.~\eqref{eq:score_prob}, this is equivalent to retaining all labels whose predicted probability exceeds a data-driven threshold,
\[
\Gamma_\alpha(X_{n+1})
\;=\;
\{\, y : \hat f_y(X_{n+1}) \ge 1-\hat q_{1-\alpha} \,\}.
\]

An equivalent and often useful formulation is in terms of conformal \emph{p}-values. For each candidate label $y\in\mathcal{Y}$, one defines
\begin{equation}\label{eq:py}
p_y(X_{n+1})
\;=\;
\frac{1 + \#\{ i\in\{1,\dots,n_{\mathrm{cal}}\} : s_i \ge s(X_{n+1},y)\}}
{n_{\mathrm{cal}}+1}.
\end{equation}
The prediction set at level $\alpha$ can then be written as
\[
\Gamma_\alpha(X_{n+1})
\;=\;
\{\, y\in\mathcal{Y} : p_y(X_{n+1}) > \alpha \,\}.
\]
Although the p-values across different labels are not independent, exchangeability ensures the finite-sample guarantee
\[
\mathbb{P}\bigl(Y_{n+1}\notin\Gamma_\alpha(X_{n+1})\bigr) \le \alpha,
\]
and hence the marginal coverage property in Eq.~\eqref{eq:marginal-cov}.

While the score in Eq.~\eqref{eq:score_prob} treats labels independently, more efficient prediction sets can often be obtained by exploiting the relative ordering of class probabilities. An important example is the Adaptive Prediction Set (APS) score introduced in Ref.~\cite{Romano2020}. For a given input $x$, let $\pi_x$ be a permutation of $\mathcal{Y}$ such that $\hat f_{\pi_x(1)}(x)\ge \hat f_{\pi_x(2)}(x)\ge\cdots\ge \hat f_{\pi_x(K)}(x)$. For a label $y\in\mathcal{Y}$, define its rank $r_x(y)$ by $\pi_x(r_x(y))=y$. The APS non-conformity score is then defined as the cumulative probability mass up to the rank of $y$,
\begin{equation}\label{eq:aps}
s_{\mathrm{APS}}(x,y) \;=\; \sum_{k=1}^{r_x(y)} \hat f_{\pi_x(k)}(x).
\end{equation}
Intuitively, $s_{\mathrm{APS}}(x,y)$ measures how much total probability mass must be accumulated before the label $y$ is included when labels are ranked by confidence.
Using $s_{\mathrm{APS}}$ in place of Eq.~\eqref{eq:score_prob} within the same split conformal or p-value framework yields prediction sets that satisfy the identical marginal coverage guarantee, while typically producing smaller and more adaptive label sets in multiclass problems. As with simpler scores, APS requires no assumptions beyond exchangeability for validity; improvements arise solely from a more informative ranking of labels and thus affect efficiency rather than coverage.

Overall, conformal classification provides a mechanism for uncertainty quantification in discrete prediction problems. By producing label sets with guaranteed marginal coverage and allowing flexible, model-dependent nonconformity scores, it offers a robust alternative to point predictions in settings where ambiguity, class overlap, or model misspecification are intrinsic, as is often the case in high-energy physics classification tasks.

In the following, we extend the study to classification problems in HEP where sec.~\ref{sec:class_binary} delves into a binary classification example investigated for various available models on {\it Top Quark Tagging Dataset}. In sec.~\ref{sec:class_multi}, we extend this implementation to multiclass classification for the {\it JetClass} dataset.

\subsection{Binary classification}
\label{sec:class_binary}

For our study of binary classification with conformal prediction, we consider three representative models, namely Particle Flow Networks (PFN)~\cite{Komiske:2019aa}, Minimal Basis for N-subjettiness (MB8S)~\cite{Moore:2018lsr}, where we use $N=8$ and Particle Flow Interaction Network (PFIN)~\cite{Khot:2022aky}. Detailed summaries of these networks are provided in Appendix~\ref{app:classification-spec}.

\begin{figure}[ht]
    \centering
    \subfigure[ROC curves\label{fig:binary_roc}]{\includegraphics[width=.485\textwidth]{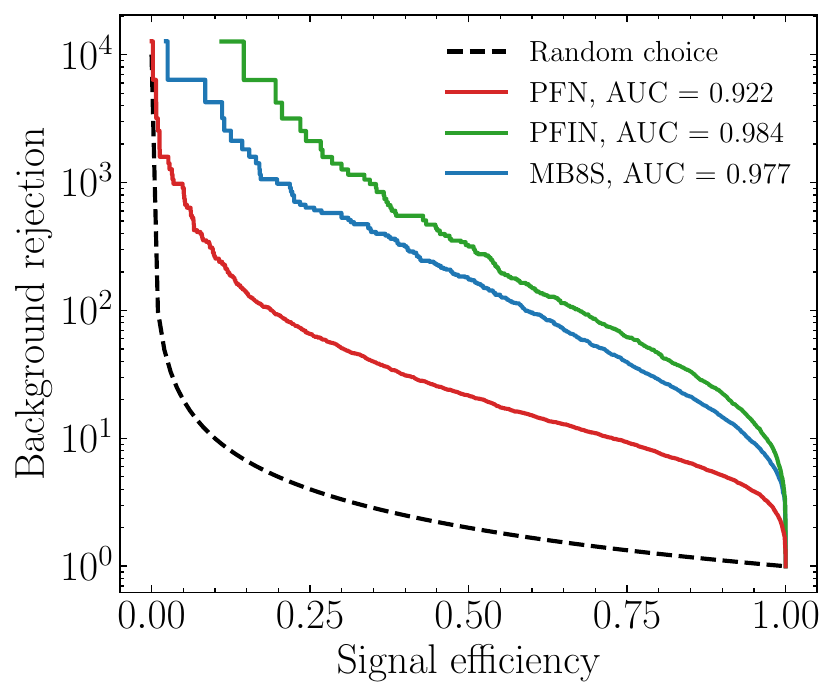}}\quad
    \subfigure[Emprical coverage and average size of the prediction sets.\label{fig:binary_cov}]{\includegraphics[width=.485\textwidth]{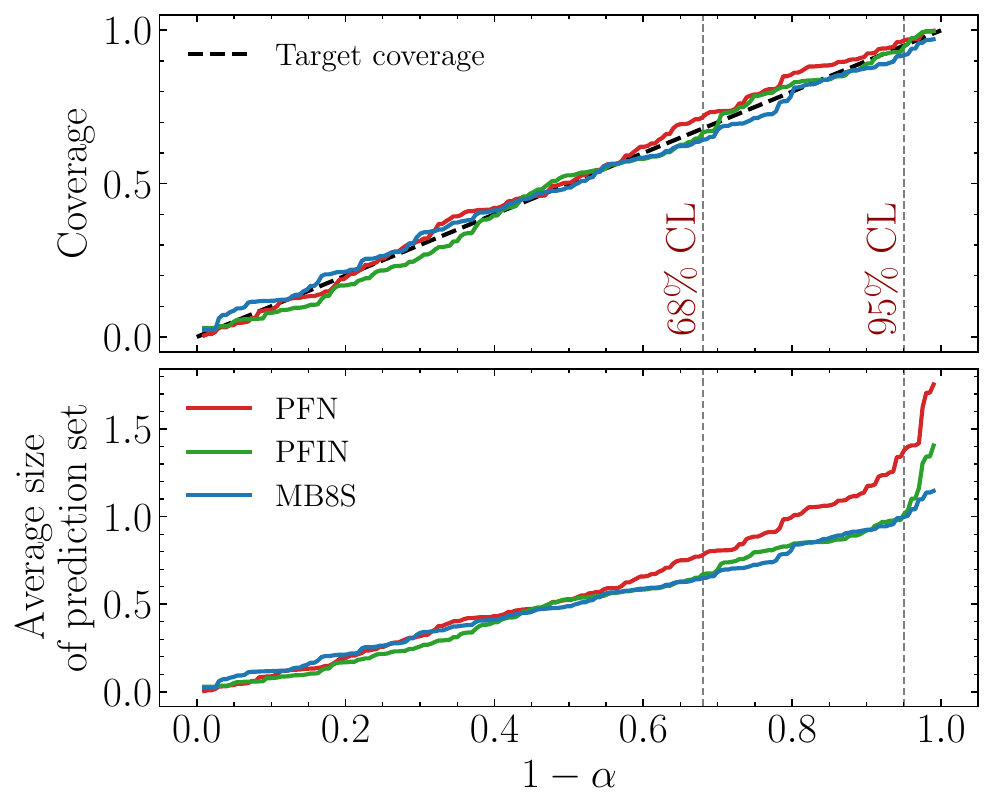}}
    \caption{Panel \textbf{(a)} shows ROC for PFN (Particle Flow Network), a PFN variant (PFIN; see the text for architecture), and Minimal Basis for 8-subjettiness (MB8S). The top of panel (b) shows empirical coverage on the test sample as a function of nominal coverage $1-\alpha$ for the three base classifiers (PFN, PFIN, MB8S), where bottom panel shows mean prediction-set size $ \mathbb{E}[\,|\Gamma_\alpha(x)|\,]$ versus $1-\alpha$. }
    \label{fig:binary}
\end{figure}

We used each model as presented in their respective publications and trained them on the top quark tagging dataset for 100 epochs.\footnote{Note that the goal of this study is not to exhaust the potential of each model; we refer the reader to the respective studies for the best performance results.} After training, the PFN, PFIN, and MB8S models achieved $85\%$, $93\%$, and $92\%$ accuracy on the validation data. Figure~\ref{fig:binary_roc} shows the ROC of the three base classifiers on the test set before conformalisation. The red, blue, and green curves represent PFN, PFIN, and MB8S, respectively, compared with a random choice indicated by the black dashed line. Each model achieved an AUC above 0.9. However, ROC curves alone provide no information about calibration or uncertainty. To derive prediction sets, we set aside 0.5\% of the test set as a calibration set, which is used exclusively for network calibration. For each event, we use the nonconformity score in Eq.~\eqref{eq:score_prob}. 

Figure~\ref{fig:binary_cov} presents the empirical coverage (top panel) and the average size of the prediction sets (bottom panel) as a function of the target coverage level $1-\alpha$, where typical confidence levels at 68\% and 95\% are represented with vertical dashed lines. The red, green, and blue curves correspond to the PFN, PFIN, and MB8S networks, respectively. As expected, all three models achieve coverage closely aligned with the nominal target values, demonstrating that the conformal procedure successfully enforces marginal validity across the full range of confidence levels.

The lower panel, on the other hand, highlights clear differences in efficiency. While the PFIN and MB8S models maintain relatively compact prediction sets over the entire interval, the PFN exhibits noticeably larger sets, particularly once the target confidence level exceeds 95\%. This behaviour is consistent with the ROC performance shown in Fig.~\ref{fig:binary_roc}: the weaker discriminating power of the PFN leads to higher uncertainty in its output scores, which the conformal method translates into broader prediction sets to preserve coverage. In contrast, the stronger, more stable predictions of PFIN and MB8S yield substantially tighter sets, reflecting their improved calibration and discriminative ability.

\begin{figure}[ht]
    \centering
    \begin{minipage}{0.5\textwidth}
        \includegraphics[width=\linewidth]{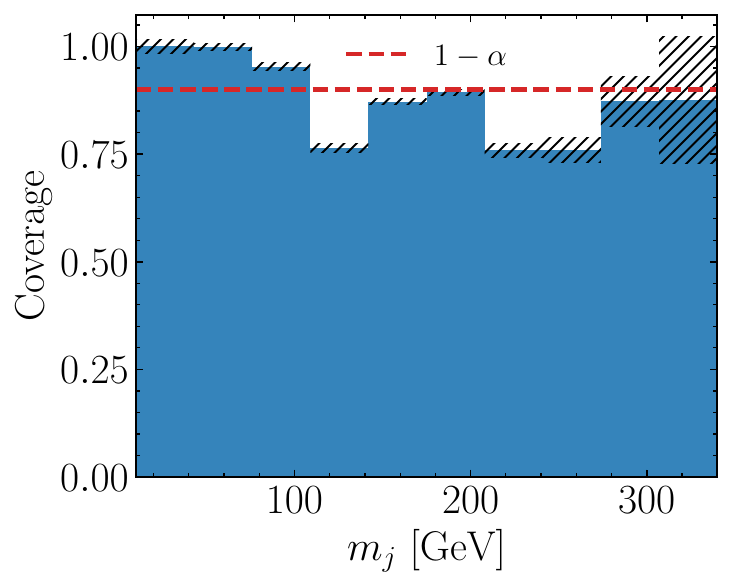}
    \end{minipage}%
    \hfill
    \begin{minipage}{0.45\textwidth}
        \caption{Conditional coverage by jet mass for PFN at $\alpha=0.1$ using the score $S(x,y)=1-p_\theta(y\mid x)$. The red dashed line shows the nominal target $1-\alpha=0.9$ and hatched caps indicate $90\%$ binomial confidence intervals per bin. 
        }
    \label{fig:binary_condcover}
    \end{minipage}
\end{figure}

Furthermore, Fig.~\ref{fig:binary_condcover} examines the empirical coverage of the conformal classifier as a function of the jet mass $m_j$ for the PFN model at a nominal confidence level of $90\%$. By construction, the conformal prediction sets achieve the desired marginal coverage when averaged over the full data distribution. However, when conditioning on a specific kinematic variable such as $m_j$, the observed coverage can deviate from the target value $1-\alpha=0.9$, as reflected by the histogram. The hatched regions indicate binomial confidence intervals, capturing the statistical uncertainty associated with the finite number of events in each bin.

The figure highlights an important and well-known property of conformal prediction: while marginal coverage is guaranteed distribution-free, conditional coverage with respect to individual features is not guaranteed in general. Deviations from the nominal level, therefore, do not signal a failure of the method, but rather reflect variations in model performance and uncertainty calibration across different regions of phase space. In particular, systematic under-coverage (over-coverage) in specific $m_j$ ranges indicates regions where the underlying classifier tends to be overconfident (underconfident), potentially due to limited training statistics, reduced feature discriminative power, or intrinsic ambiguities in jet substructure at those masses.

From a practical perspective, this behaviour provides valuable diagnostic information. Conditional coverage profiles can be used to identify kinematic regions where uncertainty estimates may be less reliable and where further mitigation strategies, such as localised conformal methods, stratified (Mondrian) conformal prediction, or targeted model improvements, may be warranted. In this sense, the observed deviations underscore both the robustness and the interpretability of conformal methods: they preserve rigorous global guarantees while transparently exposing residual structure in the learned uncertainties that is directly tied to the observables.

For the additional results, comparison between different score functions and effects of conformalisation on the confusion matrix, we refer the reader to Appendix~\ref{app:additional-binary}.

\subsection{Multi-class classification}
\label{sec:class_multi}

We now extend the analysis to multi-class classification, where each event can belong to one of several jet categories. 
For this task, we employ the publicly available \emph{Omnilearn} model~\cite{Mikuni:2024qsr}, a foundational model trained on multiple jet types. A detailed summary of Omnilearn is provided in Appendix~\ref{app:classification-spec}.

Following the default construction presented in Ref.~\cite{Mikuni:2024qsr}, we train the multiclass classifier on the JetClass dataset for 20 epochs. The resulting model achieves a validation accuracy of $77\%$, with the lowest area under the receiver operating characteristic curve on the test set being $0.94$ for the $H\to c\bar c$ process\footnote{We refer the reader to the original publication for a detailed optimisation study and state-of-the-art performance benchmarks.}. These metrics indicate strong discriminative power across ten jet categories considered, as assessed by conventional classification measures.

For the conformal calibration stage, we set aside a calibration sample of 500 randomly selected events, without enforcing equal class representation. This choice intentionally reflects a realistic, constrained calibration scenario in which both the overall sample size and the per-class statistics are limited. As the performance of conformal prediction depends on the empirical distribution of calibration scores, the size and class composition of the calibration set play a central role in determining the sharpness of the resulting prediction sets.

\begin{figure}[ht]
    \centering
    \begin{minipage}{0.55\textwidth}
        \includegraphics[width=\linewidth]{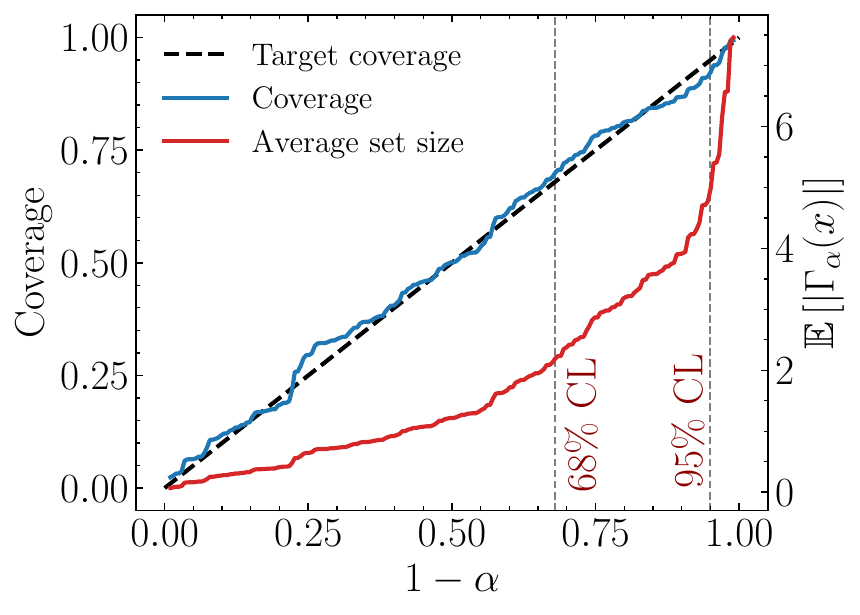}
    \end{minipage}%
    \hfill
    \begin{minipage}{0.4\textwidth}
        \caption{Global conformal prediction performance for multi-class classification using the Omnilearn model. The blue curve shows empirical test-set coverage as a function of the nominal target $1-\alpha$, and the dashed black line shows the target coverage. The red curve shows the corresponding average prediction-set size $\mathbb{E}[|\Gamma_\alpha(x)|]$. Vertical dashed lines mark two representative confidence levels at $68\%$ and $95\%$ CL.}
    \label{fig:multi_cov_size}
    \end{minipage}
\end{figure}

Given the multiclass nature of the problem and the presence of ten competing processes in JetClass, we adopt the Adaptive Prediction Set (APS) score for calibration as defined in Eq.~\eqref{eq:aps}. APS is designed to minimise prediction set sizes by exploiting the ranked structure of the classifier outputs, and is therefore well suited to multiclass settings. Figure~\ref{fig:multi_cov_size} shows the resulting empirical coverage (blue) and average prediction set size (red) as functions of the nominal coverage level $1-\alpha$. As expected, the coverage remains close to the target value across the full range of confidence levels, confirming the validity of the conformal construction. At the same time, the average prediction set size increases rapidly as higher confidence levels are demanded, reaching approximately five labels at $95\%$ confidence. This behaviour reveals that despite good performance on accuracy and ROC-based metrics, the classifier exhibits substantial ambiguity when required to make highly confident assignments, reflecting significant class overlap in certain regions of feature space.

\begin{figure}[ht]
    \centering
    \begin{minipage}{0.55\textwidth}
        \includegraphics[width=\linewidth]{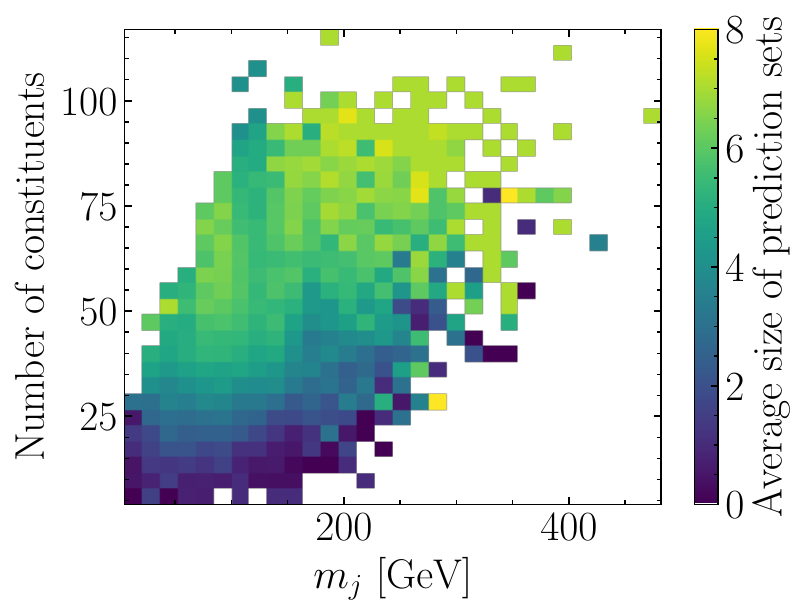}
    \end{minipage}%
    \hfill
    \begin{minipage}{0.4\textwidth}
        \caption{Average prediction–set size $\mathbb{E}[|\Gamma_\alpha(x)|]$ for multi-class conformal classification using the Omnilearn model, shown in the $(m_j,\,N_{\mathrm{const}})$ plane at $1-\alpha=0.9$. Colour indicates the mean number of labels retained in the conformal prediction set for events in each kinematic bin; white bins have no statistics.}
    \label{fig:multimeanpred}
    \end{minipage}
\end{figure}

To better understand the origin and structure of this uncertainty, we examine how the average size of the prediction set varies across phase space. Figure~\ref{fig:multimeanpred} presents a two-dimensional histogram in jet mass and number of jet constituents, where the colour scale represents the mean size of the conformal prediction set at $90\%$ confidence level. The figure shows that the model is most certain for low-mass jets ($m_j \lesssim 200~\mathrm{GeV}$) with relatively few constituents, where the prediction sets are typically small. As the jet mass increases, the model gets more confident with larger numbers of constituents. In particular, regions with large constituent multiplicities (around 75 constituents) exhibit pronounced uncertainty, suggesting that the learned representation struggles to disentangle the underlying processes in this regime. Such phase-space-resolved diagnostics are highly informative, as they directly point to regions where additional feature engineering, architectural modifications, or targeted training strategies could reduce uncertainty.

\begin{figure}[ht]
    \centering
    \begin{minipage}{0.55\textwidth}
        \includegraphics[width=\linewidth]{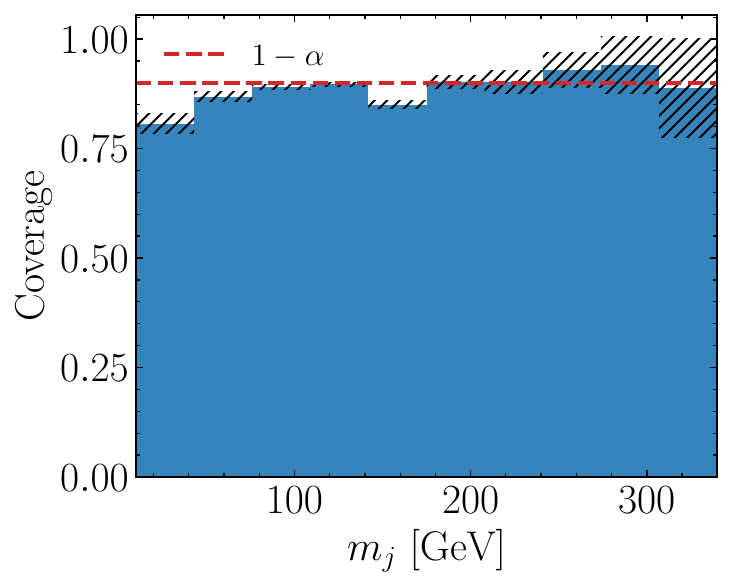}
    \end{minipage}%
    \hfill
    \begin{minipage}{0.4\textwidth}
        \caption{Conditional coverage for multi-class conformal prediction using the Omnilearn classifier at $\alpha=0.10$, shown as a function of jet mass $m_j$. The red dashed line marks the nominal target coverage $1-\alpha=0.9$, and hatched regions denote the corresponding $90\%$ binomial confidence intervals per bin. }
    \label{fig:multicondcov}
    \end{minipage}
\end{figure}

Finally, Fig.~\ref{fig:multicondcov} investigates the empirical coverage conditional on jet mass at the same nominal confidence level. Consistent with this expectation, the figure shows moderate variations in coverage across different $m_j$ bins, with binomial error bars quantifying the statistical uncertainty in each bin. These deviations illustrate the well-known distinction between marginal and conditional validity in conformal inference and should not be interpreted as a breakdown of the method. Instead, from a physics perspective, they highlight regions of mass where the classifier tends to be systematically over- or under-confident. Such information can be used to guide targeted improvements to the model.

These results demonstrate that conformal prediction provides a robust and interpretable framework for uncertainty quantification in multiclass jet classification. By transforming classifier outputs into prediction sets with rigorous statistical guarantees, CP exposes aspects of model uncertainty that are invisible to traditional performance metrics alone. In this way, conformal methods complement accuracy- and ROC-based evaluations, enabling a more nuanced assessment of classifier reliability across the whole kinematic landscape.

\section{Conformal calibration for anomaly detection}
\label{sec:anomaly}

Using the same formalism described for regression and classification problems, one can extend CP to anomaly detection. For this study, we have chosen three network architectures proposed in Ref.~\cite{Liu:2023djx}, namely Deepset-Set, Transformer-Clip and Transformer-set variational autoencoders. Detailed summaries of these models are provided in the Appendix~\ref{app:anomaly}.  

For the anomaly detection study, we utilise the pretrained autoencoder-based networks from the original analysis, which were trained exclusively on QCD jets. As a result, the learned representations capture the typical structure of QCD radiation patterns, while any deviation from these patterns may be interpreted as anomalous. The input features comprise the momenta and energies of jet constituents expressed in cylindrical coordinates relative to the jet axis, together with discrete particle-type information for eight categories (charged and neutral hadrons, photons, electrons, and muons). 

The nonconformity score is defined as the Chamfer loss between the input jet and its autoencoder reconstruction, computed under the constraint that only particles with matching discrete labels contribute to the distance. Concretely, the score is obtained by summing the minimum squared distances between reconstructed and original particles of the same type, ensuring that discrepancies in both kinematic structure and particle content are penalised. As in previous sections, we reserve 500 events from the test sample for calibration; these events consist exclusively of QCD jets and are disjoint from the training set. The conformal threshold $\hat q_{1-\alpha}$ is determined from the empirical distribution of calibration scores via Eq.~\eqref{eq:score-quantile}. In contrast to classification, events with scores exceeding this threshold are labelled as anomalous, yielding a decision rule that guarantees control of the false positive rate for QCD jets at level $\alpha$ under the exchangeability assumption.

For anomaly detection, it is useful to characterise performance in terms of the \emph{signal efficiency} under conformal prediction. Given a nonconformity score $s(X)$ and a conformal threshold $\hat q_{1-\alpha}$ obtained from background-only calibration via Eq.~\eqref{eq:score-quantile}, an event is classified as anomalous if $s(X) > \hat q_{1-\alpha}$. The signal efficiency at confidence level $1-\alpha$ is then defined as
\begin{equation}
\varepsilon_{\mathrm{sig}}(1-\alpha)
\;=\;
\mathbb{P}_{X \sim P_{\mathrm{sig}}}
\bigl( s(X) > \hat q_{1-\alpha} \bigr),
\end{equation}
where $P_{\mathrm{sig}}$ denotes the distribution of signal events. 

\begin{figure}[ht]
    \centering
    \subfigure[Background coverage\label{fig:anomaly-bkg-cov}]{\includegraphics[width=0.45\linewidth]{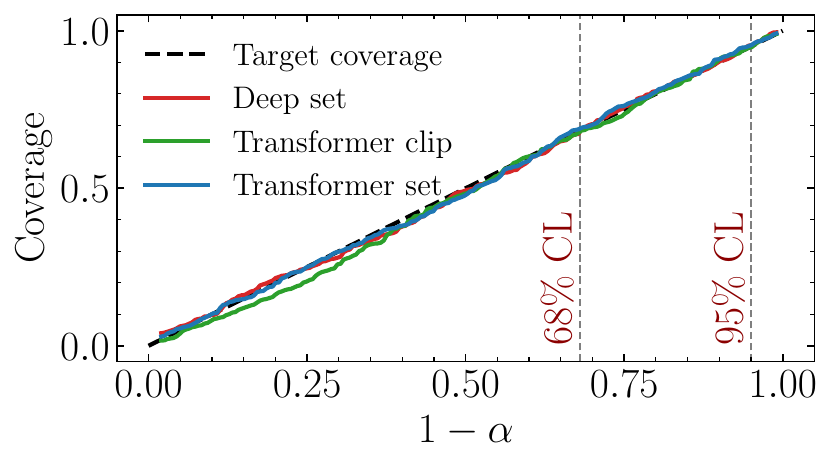}}\qquad
    \subfigure[Signal efficiency vs $1-\alpha$\label{fig:anomaly-sig-eff}]{\includegraphics[width=0.45\linewidth]{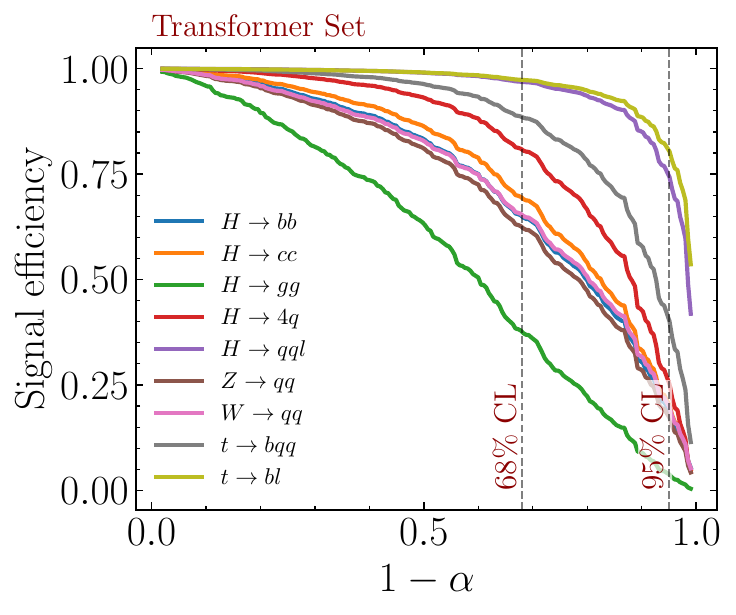}}
    \caption{Anomaly detection with conformal calibration. 
        (a) Empirical background coverage as a function of nominal confidence level $1-\alpha$ for conformal $p$-values constructed from the anomaly score. 
        The dashed diagonal indicates ideal coverage. 
        (b) Signal efficiency as a function of $1-\alpha$ for a TransformerSet–VAE anomaly model, defined as the fraction of signal events with calibrated $p(x)<\alpha$. 
        While conformal prediction enforces exact background control by construction, the signal curves illustrate how different processes are separated by the calibrated anomaly score.
    }
    \label{fig:cov-sigeff}
\end{figure}

Figure~\ref{fig:anomaly-bkg-cov} summarises the background (QCD) coverage as a function of the nominal confidence level $1-\alpha$ for the three considered architectures: DeepSets (red), Transformer-CLIP (green), and Transformer Set (blue). All models closely track the target coverage, shown as a dashed black line, thereby validating the conformal calibration procedure independently of the underlying architecture. Figure~\ref{fig:anomaly-sig-eff} complements this result by showing the corresponding signal efficiency for the Transformer Set autoencoder across different signal processes. A clear hierarchy emerges: the lowest signal efficiency is observed for the $H\to gg$ process, while processes containing leptons in the final state, such as $t\to b\ell$ and $H\to q q \ell$, maintain efficiencies of order $75\%$ even at $95\%$ confidence. This behaviour is physically intuitive, as the presence of leptons constitutes a strong deviation from the QCD training distribution. The same qualitative pattern is observed across all model architectures (see Fig.~\ref{fig:app-sig-alpha}), indicating that the dominant sources of anomaly sensitivity are shared and are driven primarily by particle content.

\begin{figure}[ht]
    \centering
    \subfigure[Transverse momentum\label{fig:anomaly-cov-mom}]{\includegraphics[width=.49\linewidth]{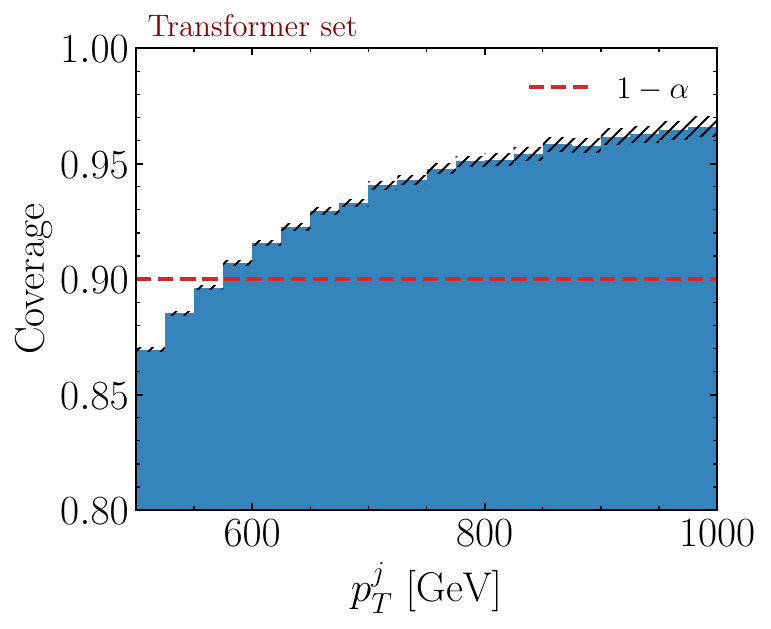}}
    \subfigure[Soft-drop mass\label{fig:anomaly-cov-mass}]{\includegraphics[width=.49\linewidth]{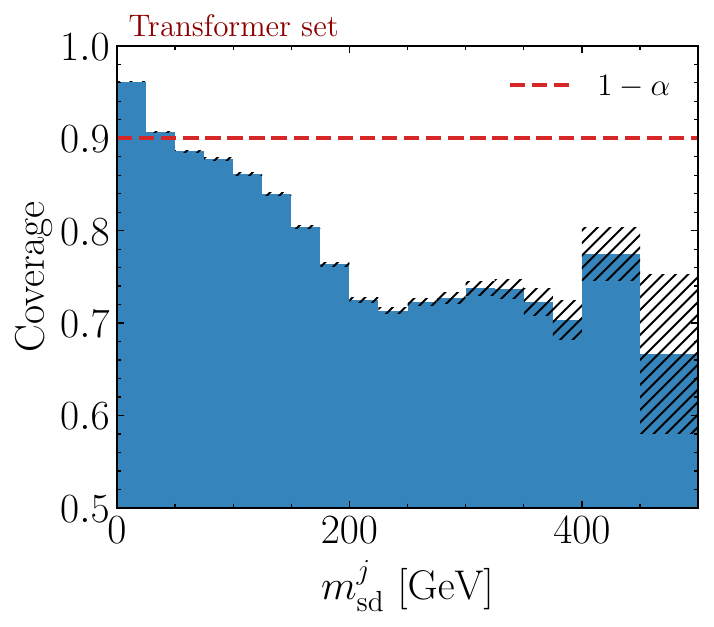}}
        \caption{Conditional background coverage of conformal $p$-values for the TransformerSet anomaly model. The empirical coverage is shown as a function of jet transverse momentum $p_T$ (left) and soft-drop jet mass $m_{\mathrm{sd}}$ (right). Bars indicate the fraction of background events in each kinematic bin with calibrated $p(x)>\alpha$, and the dashed line marks the nominal target coverage $1-\alpha$. Deviations across bins reflect the fact that split conformal prediction guarantees marginal, but not conditional, coverage.
    }
    \label{fig:cov-anomaly-transformerset}
\end{figure}

Following the same procedure as in the classification case, Fig.~\ref{fig:cov-anomaly-transformerset} presents the empirical background (QCD) coverage as a function of transverse momentum (panel a) and soft-drop jet mass (panel b), with binomial confidence intervals indicated by hatched bands. By construction, the overall marginal coverage is guaranteed; however, conditioning on specific kinematic variables may lead to local deviations from the target coverage. For the transverse momentum distribution, we observe a smooth and gradual increase in coverage with jet energy, suggesting that the autoencoder reconstruction becomes increasingly conservative at higher momenta. In contrast, the soft-drop mass distribution exhibits overcoverage predominantly in the lowest-mass bin. These trends reflect variations in how well the learned QCD representation captures different kinematic regimes, and illustrate once again the distinction between marginal and conditional coverage in conformal inference.

\begin{figure}[h]
    \centering
    \includegraphics[width=.95\linewidth]{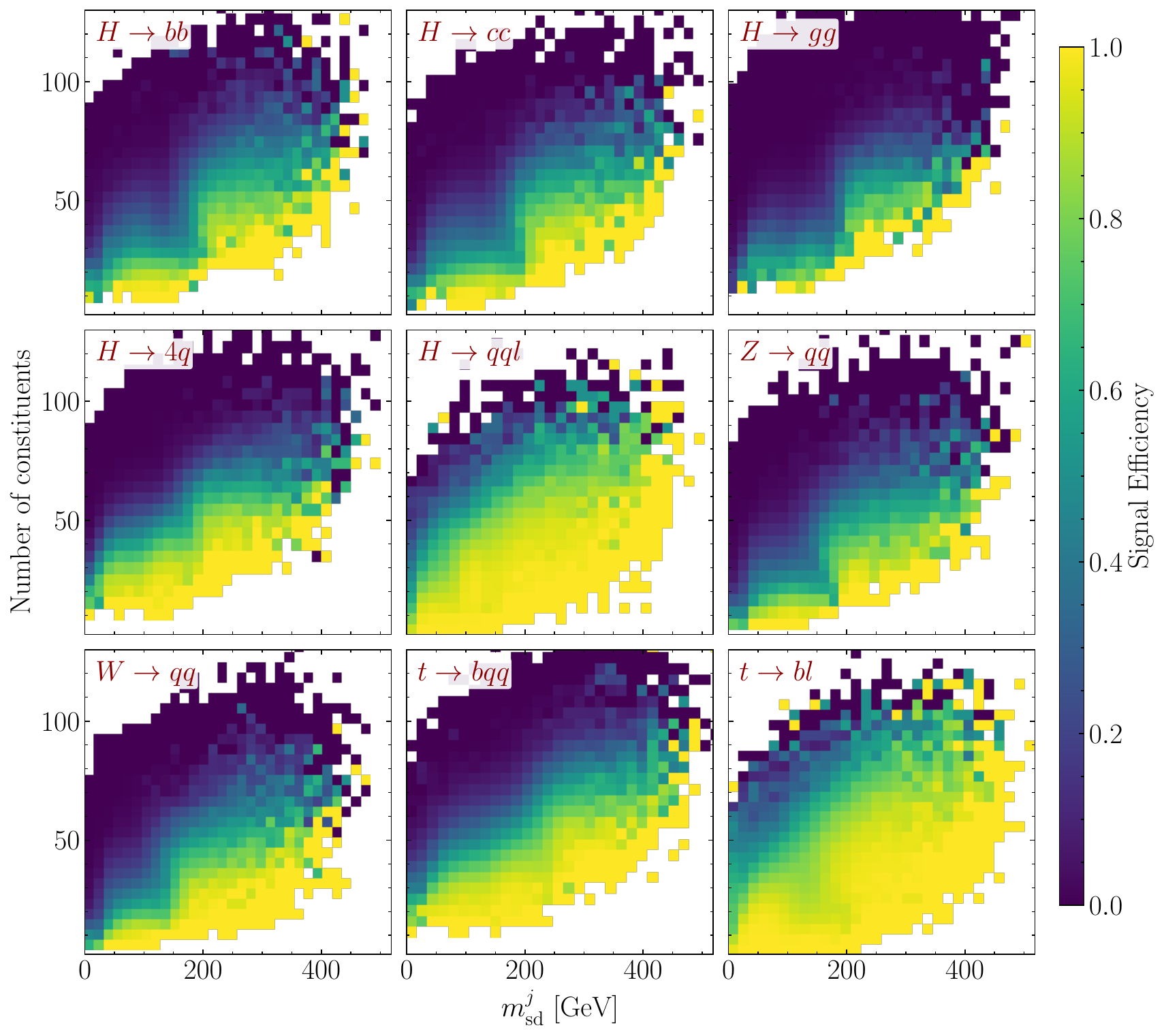}
    \caption{Kinematic dependence of anomaly-detection sensitivity for the TransformerSet–VAE model at $1-\alpha=0.9$. The colour scale shows the signal efficiency, defined as the fraction of signal events with calibrated $p(x)<\alpha$, in two-dimensional bins of jet mass $m_j$ and number of constituents.}
    \label{fig:kin-anomaly-transformerset}
\end{figure}

The interpretability afforded by conformal prediction is further illustrated in Fig.~\ref{fig:kin-anomaly-transformerset}, which maps the signal efficiency at $90\%$ confidence onto the two-dimensional phase space spanned by soft-drop mass and the number of jet constituents for the Transformer Set autoencoder. Each bin is coloured according to the average signal efficiency in that region. This representation reveals where in phase space the anomaly detector is most and least sensitive. For the $H\to gg$ signal, sensitivity is concentrated in the region of high jet mass and relatively low constituent multiplicity. In contrast, signals with leptonic final states exhibit a more uniform efficiency across mass, but lose sensitivity at high constituent multiplicities, where the jet substructure increasingly resembles that of QCD. Across all signal models, a common trend emerges: the anomaly detector systematically loses sensitivity in regions with many constituents, reflecting that dense particle environments provide more opportunities for the autoencoder to reconstruct the jet as QCD-like.

Crucially, these phase-space-resolved efficiency maps go beyond being a simple performance metric and provide direct insight into the model's learned notion of {\it normality}. By coupling conformal calibration with physically interpretable nonconformity scores, the approach not only delivers statistically controlled anomaly detection but also enables a detailed diagnosis of where and why the model succeeds or fails. This level of interpretability is particularly valuable in high-energy physics applications, as it allows model behaviour to be scrutinised and improved in targeted regions of kinematic phase space, and facilitates a principled assessment of discovery sensitivity under well-defined statistical guarantees.

The complementary results for other models and their comparisons are provided in Appendix~\ref{app:anomaly-additional}.

\section{Conformal calibration of generative models}
\label{sec:generative}

Generative models are increasingly important in collider physics, with applications ranging from fast simulation and detector emulation to density estimation, reweighting, and likelihood-free inference. Normalising flows, variational autoencoders, diffusion models, and other architectures can approximate complex multidimensional probability distributions and generate synthetic events that closely resemble real data. However, these models typically lack statistical reliability guarantees: a generative model may reproduce some regions of phase space faithfully while exhibiting distortions or mode collapse in others.  
Moreover, the quantities these models yield, such as negative log-likelihoods, reconstruction discrepancies, or latent-space distances, have no universal meaning without calibration. A score that appears large in one kinematic region may be typical in another, making it difficult to compare, interpret, or threshold such outputs in a principled way.

We consider the application of conformal prediction to {\it generative modelling}, where the goal is not to predict labels or label sets, but rather to assess and calibrate the fidelity of samples drawn from a learned probability distribution. As a concrete example, we employ the normalising flow (NF) model introduced in Ref.~\cite{Araz:2025ezp} to represent the probability density of ATLAS OmniFold data for the process $pp \to Z(\to \mu^+\mu^-) + \mathrm{jets}$.

While the original dataset consists of 24 observables, the NF was trained on a reduced five-dimensional phase space that captures the dominant kinematic structure of the underlying \(2\to3\) scattering. These observables are the transverse momentum and rapidity of the dimuon system \((p_T^{\mu\mu}, y^{\mu\mu})\), the transverse momentum of the leading muon \(p_T^{\mu_1}\), the pseudorapidity difference \(\Delta\eta_{\mu\mu}\), and the azimuthal angle difference \(\Delta\phi_{\mu\mu}\).

Prior to training, all observables were shifted and rescaled to lie in the unit interval \([0,1]\). For transverse momentum variables, an additional logarithmic transformation was applied before rescaling, improving numerical stability and enabling the NF to resolve both the bulk and tail regions of the distribution more efficiently.

The likelihood model was implemented as an eight-layer normalising flow using Masked Autoregressive Flow (MAF)~\cite{Papamakarios:2017tec} with Rational Quadratic Spline (RQS)~\cite{durkan2019neuralsplineflows} transformations. Throughout this study, we used a publicly available pretrained model from the Zenodo repository~\cite{anja_beck_2025_14902619}. The quality of the learned density was assessed using standard goodness-of-fit tests on held-out data. The Kolmogorov-Smirnov test yields a \(p\)-value of 66.5\%, while a binned \(\chi^2\) test gives a \(p\)-value of 50.7\%, indicating that the NF provides an adequate description of the target distribution at the level of traditional global tests.

While such tests probe overall agreement, they do not provide a \emph{distribution-free, finite-sample guarantee} on the typicality of individual events. To address this, we apply conformal prediction to the generative model following the probabilistic conformal framework of Ref.~\cite{wang2022probabilisticconformalpredictionusing}. In this setting, the generative model defines a likelihood \(p_\theta(x)\), and conformalisation proceeds by defining a nonconformity score
\begin{equation}
s(x) \;=\; -\log p_\theta(x),
\end{equation}
so that events assigned a low likelihood by the model are deemed less conforming to the learned distribution.

Using a calibration set \(\{X_i\}_{i=1}^{n_{\mathrm{cal}}}\) of 500 events drawn from the test sample, independent of the training data, we compute the corresponding scores \(s_i = s(X_i)\). The conformal threshold \(\hat q_{1-\alpha}\) is then obtained via Eq.~\eqref{eq:score-quantile}. In contrast to classification or regression, where conformal prediction produces sets of outputs, here the threshold induces a \emph{typicality region} in data space:
\begin{equation}
\mathcal{T}_{1-\alpha}
\;=\;
\{\, x : s(x) \le \hat q_{1-\alpha} \,\}.
\end{equation}
By construction, and under exchangeability, this region satisfies
\[
\mathbb{P}_{X \sim P_{\mathrm{data}}}(X \in \mathcal{T}_{1-\alpha}) \ge 1-\alpha,
\]
providing a finite-sample coverage guarantee for samples drawn from the true data distribution.

\begin{figure}[h]
    \centering
    \subfigure[coverage curve for conformal p-values\label{fig:generative-cov-alpha}]{\includegraphics[width=.47\linewidth]{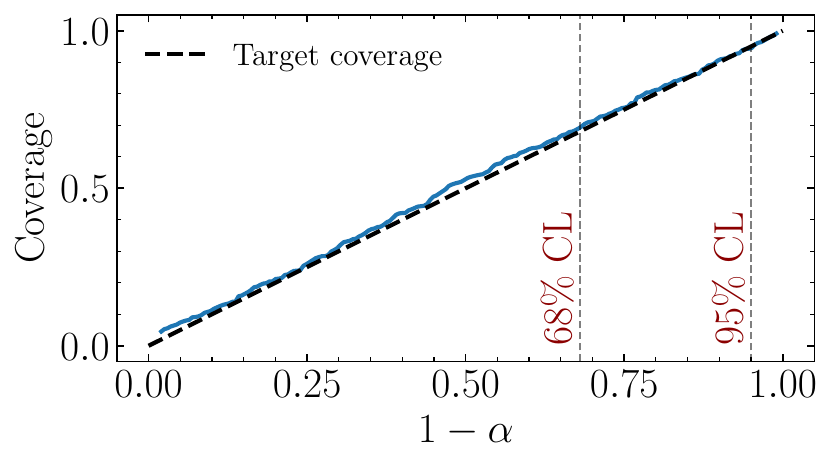}}\qquad
    \subfigure[coverage against $p_T^{\mu\mu}$\label{fig:generative-cov-pt}]{\includegraphics[width=.47\linewidth]{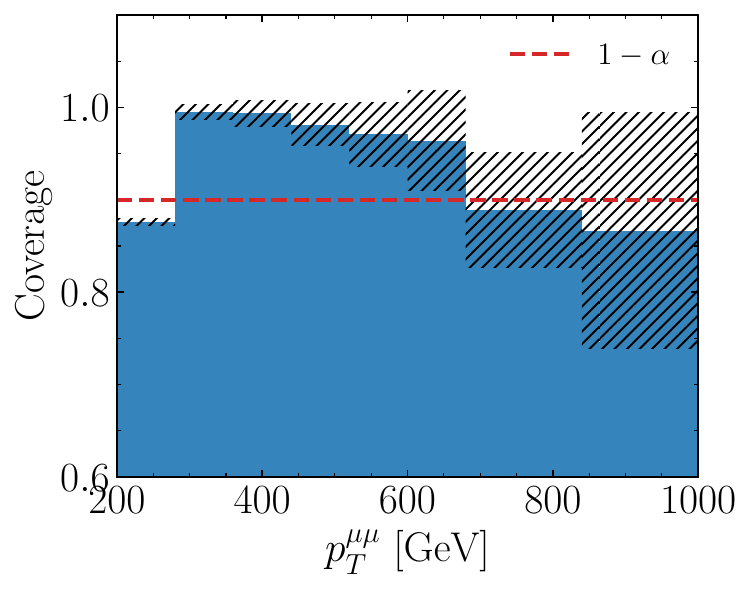}}
    \caption{(a) Global coverage curve for conformal p-values extracted from the generative-model discrepancy score. (b) Conditional coverage as a function of the dimuon transverse momentum $p^{\mu\mu}_T$ at $1-\alpha=0.90$. Bars show the fraction of events in each $p_T$ bin with p-value above $\alpha$, with binomial uncertainty bands indicated by the hatched regions.}
    \label{fig:generative-cov}
\end{figure}

To validate the conformalisation, we generate 20,000 events from the NF and evaluate their scores against the calibrated threshold. Figure~\ref{fig:generative-cov-alpha} shows the empirical coverage as a function of \(1-\alpha\), demonstrating close agreement with the target coverage across the full range of miscoverage levels. This confirms that the conformal procedure successfully calibrates the generative model, independent of the accuracy of the underlying likelihood approximation.

Figure~\ref{fig:generative-cov-pt} further probes \emph{conditional coverage} as a function of the dimuon transverse momentum \(p_T^{\mu\mu}\) at 90\% confidence level. The conditional coverage exhibits non-uniform behaviour: the mid-range of the spectrum shows mild over-coverage, whereas the high-\(p_T\) tail undercovers, albeit with large binomial uncertainties due to limited statistics. This behaviour is consistent with known limitations of generative models in sparsely populated regions of phase space.

\begin{figure}[h]
    \centering
    \includegraphics[width=.95\linewidth]{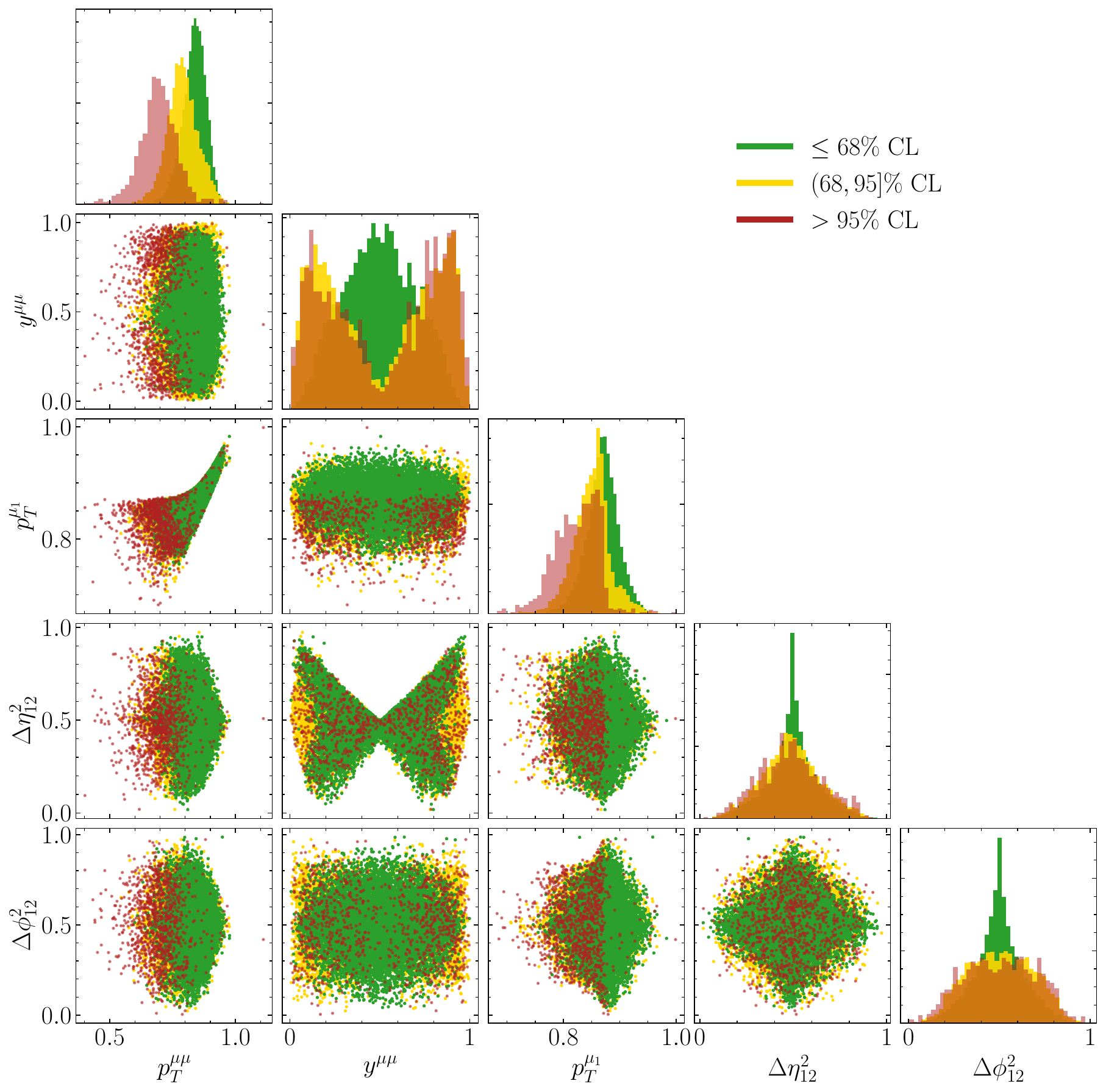}
    \caption{Conformal p-values visualised across several pairs of kinematic variables for a representative generative model.  A standardised version of the model discrepancy score is used to avoid the influence of heavy tails. Colours indicate p-value categories: $p(x)\le 0.68$ (green), $0.68 < p(x) \le 0.95$ (yellow), and $p(x) > 0.95$ (red).}
    \label{fig:generative}
\end{figure}

The interpretability gains afforded by conformalisation are further illustrated in Fig.~\ref{fig:generative}, which presents a two-dimensional corner plot of the calibrated generative distribution. Each point is coloured according to whether it lies within the 68\%, (68, 95]\%, or \(>95\%\) conformal typicality regions. The diagonal panels show the corresponding marginal densities. Owing to the rarity of events generated far from the bulk of the distribution, the figure is shown in a standardised coordinate frame for clarity.

Several physically meaningful features emerge. Events at the edges of the pseudorapidity distribution are predominantly assigned to higher conformal levels, indicating that they are statistically rare but still consistent with the learned distribution. Similarly, different conformal regions peak at different values of \(p_T^{\mu\mu}\), revealing how distinct kinematic regimes contribute unevenly to the overall likelihood structure. These effects are largely invisible to traditional one-dimensional goodness-of-fit tests but are made explicit by the conformal framework.

Overall, conformalised generative modelling provides a principled, distribution-free layer of uncertainty quantification on top of likelihood-based generative models. It enables sampling procedures with guaranteed coverage, facilitates detailed diagnostics of phase-space regions where the model is least reliable, and offers an interpretable notion of typicality that complements classical statistical tests. In this sense, conformal prediction transforms generative modelling from a purely approximate density-estimation task into one with rigorous statistical guarantees.

\section{Discussion and outlook}
\label{sec:conclusion}

We have examined conformal prediction as a general statistical framework for uncertainty quantification across a range of machine-learning tasks relevant to high-energy physics, including regression, classification, anomaly detection, and generative modelling.
Rather than proposing new learning architectures or optimisation strategies, our focus has been on the statistical interpretation of model outputs and on enforcing reliable uncertainty statements in finite samples. 
Across all settings considered, conformal prediction provides a unifying post-processing layer that transforms arbitrary model-dependent scores into quantities with well-defined and verifiable statistical meaning.

The takehome message of our study is that conformal prediction should not be viewed as a performance-enhancing technique. 
It does not improve class separability, sharpen regression estimates, or compensate for model misspecification. 
Instead, its value lies in enforcing {\it honest uncertainty}: prediction intervals, prediction sets, and anomaly decisions produced by conformal methods satisfy coverage or false-positive guarantees by construction, independently of the underlying model architecture, distribution, or training procedure. 

Our results for regression illustrate that conformalised intervals recover correct empirical coverage even in strongly heteroscedastic settings, where naive uncertainty estimates fail. 
In classification, conformal prediction sets provide a principled alternative to confidence thresholds, enabling controlled abstention and transparent trade-offs between coverage and efficiency for both binary and multi-class problems. 
For anomaly detection and generative models, conformal calibration converts unstructured anomaly scores into calibrated $p$-values, yielding exact finite-sample control of the background false-positive rate and enabling meaningful comparisons across models, phase-space regions, and physics processes. 
Taken together, these examples demonstrate that conformal prediction supplies the missing statistical layer that connects modern machine learning to the decision-theoretic requirements of experimental analyses.

From a broader perspective, our findings suggest several directions for future work. 
One immediate extension is to incorporate conditional or locally adaptive conformal methods to address observed variations in conditional coverage across kinematic regions. 
Another promising avenue is the integration of conformal prediction with simulation-based inference and likelihood-free methods, where calibrated uncertainty estimates are essential but often difficult to obtain. 
Conformal techniques may also play a role in the propagation of systematic uncertainties, for example, by calibrating model outputs separately across systematic variations or nuisance-parameter ensembles.

It is important to highlight two aspects of conformal prediction that warrant further investigation and naturally define directions for future work. The first concerns the size and composition of the calibration dataset. In this study, only a relatively small fraction of the available data was used for calibration. While this choice is unlikely to be limiting for well-controlled, homogeneous datasets, it may become more consequential in settings with heterogeneous data sources, such as multiclass classification problems or analyses that combine multiple production channels. In such scenarios, stratified calibration, allocating calibration samples on a per-class or per-source basis, may be necessary to ensure reliable coverage. A systematic study of the trade-off between calibration set size, data heterogeneity, and statistical efficiency would therefore be a valuable extension of the present work.

A second, closely related consideration is the choice of nonconformity score. Throughout this analysis, we employed standard nonconformity scores commonly used in the literature, which are largely agnostic to the underlying structure of specific HEP observables. Designing problem-aware nonconformity scores that exploit known physical features, correlations, or uncertainties could yield tighter predictive sets while maintaining rigorous coverage guarantees. Such score engineering may also mitigate some of the demands on the size of the calibration data. We view the joint optimisation of calibration strategies and nonconformity scores as a promising avenue for enhancing the practical impact of conformal prediction in HEP applications.

More generally, we view conformal prediction as a candidate \emph{standard tool} for uncertainty quantification in machine-learning-based analyses in high-energy physics. 
Its assumptions are minimal, its guarantees are explicit, and its implementation is model-agnostic. We therefore advocate the routine inclusion of conformal diagnostics, such as coverage curves and prediction-set size distributions, alongside traditional performance metrics, such as ROC curves or goodness-of-fit metrics. Adopting such practices would improve the interpretability and robustness of machine-learning results. It would facilitate their integration into the statistical workflows that underpin precision measurements and searches for new physics.

\addcontentsline{toc}{section}{Acknowledgment}
\section*{Acknowledgment}
JYA is supported by the Institute for Particle Physics Phenomenology Associateship Scheme and by the Royal Society under grant no. IES/R2/252139.

\addcontentsline{toc}{section}{References}
\bibliography{references}

\addcontentsline{toc}{section}{Appendix}
\appendix

\section{Classification}

In this section, we present additional material and details about the models that have been used in sections~\ref{sec:class_binary} and \ref{sec:class_multi}.

\subsection{Model Specifications}
\label{app:classification-spec}

For the application of CP to classification examples, we used four different publicly available models as summarised below;

\begin{itemize}
    \item {\bf Particle Flow Networks} (PFNs) are a class of deep learning architectures introduced in Ref.~\cite{Komiske:2019aa} that are specifically designed to process sets of particles in a way that is permutation invariant and physically meaningful for collider physics applications. In a PFN, each jet or event is represented as an unordered set of constituent particles, where each particle is described by a vector of features (such as momentum components and particle identification information). The network first applies a shared “per-particle” embedding function to each constituent, transforming particle features into a latent representation. Then it aggregates these embeddings using a symmetric pooling operation (typically a sum) to produce a fixed-length global feature vector. This aggregated representation is passed through subsequent dense layers. By construction, PFNs respect the permutation symmetry of sets and can naturally incorporate both kinematic and categorical particle information. 

    \item {\bf MB8S}~\cite{Moore:2018lsr} introduces a dense, fully connected neural network, referred to as the MBNS (Minimal Basis for $N$-subjettiness) network, designed to classify boosted jets using a compact and physically interpretable set of $N$-subjettiness observables. These observables quantify the jet's radiation pattern and multi-prong substructure by measuring how constituent momenta cluster around one, two, or three candidate subjet axes, thereby efficiently capturing the characteristic topology of hadronically decaying heavy particles such as top quarks. The MBNS network takes as input a minimal yet complete basis of these $N$-subjettiness variables, along with the jet mass, and processes them through several fully connected layers. In this study, we set $N=8$.

    \item The {\bf Particle Flow Interaction Network} (PFIN) is a machine-learning architecture proposed in Ref.~\cite{Khot:2022aky} that augments the traditional Particle Flow Network (PFN) by explicitly modelling pairwise interactions among jet constituents to improve classification performance in jet-tagging tasks. The PFIN architecture incorporates an interaction network, a graph-style module that processes latent embeddings of particle pairs, allowing the model to learn relational information between constituents in addition to individual particle features. 

    \item {\bf OmniLearn}~\cite{Mikuni:2024qsr} is a foundation model framework for jet physics that is trained on a high-dimensional multiclass jet classification task to learn a generalised representation of jet structure that can be transferred to other downstream tasks in collider physics. In the context of multiclass classification, OmniLearn uses the JetClass dataset, comprising tens of millions of jets labelled by 10 distinct classes, as its primary training objective, optimising a cross-entropy loss to distinguish among multiple jet types. The model architecture, often implemented with a Point-Edge Transformer backbone, produces a shared representation from per-particle inputs, which is then passed through a classification head that outputs probabilities over all classes.
\end{itemize}

The first three models, namely PFN, MB8S and PFIN, have been used for binary classification, and OmniLearn has been used for multiclass classification. Note that we used the models with default specifications as described in the respective publications.

\subsection{Additional results for conformal prediction in binary classification}
\label{app:additional-binary}

This appendix collects additional results and validation studies for conformal prediction in the binary classification setting. These results complement the main discussion in Section~\ref{sec:classification} by providing more detailed diagnostic views of how conformal prediction sets behave across kinematic phase space, model architectures, and operating points.

\begin{figure}
    \centering
    \subfigure[PFN\label{fig:pfn-phasespace}]{\includegraphics[width=.45\linewidth]{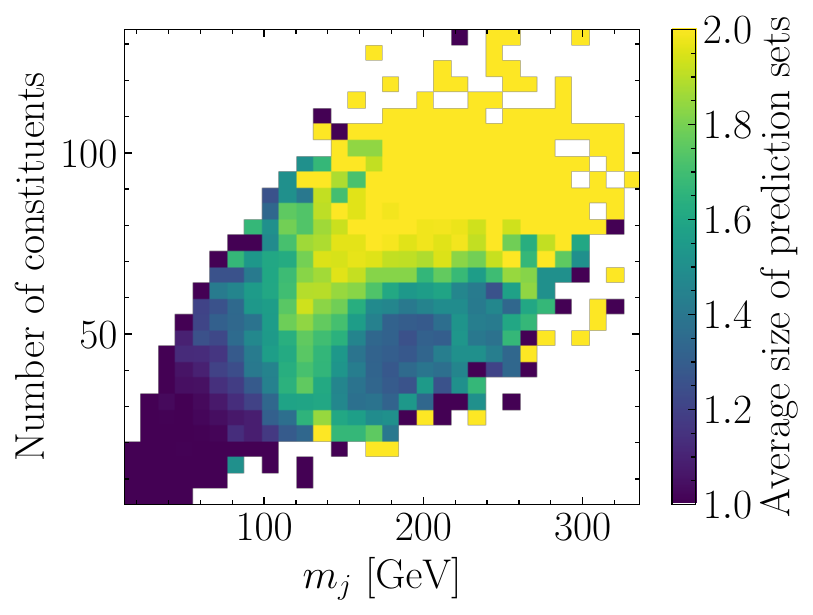}}\qquad
    \subfigure[PFIN\label{fig:pfin-phasespace}]{\includegraphics[width=.45\linewidth]{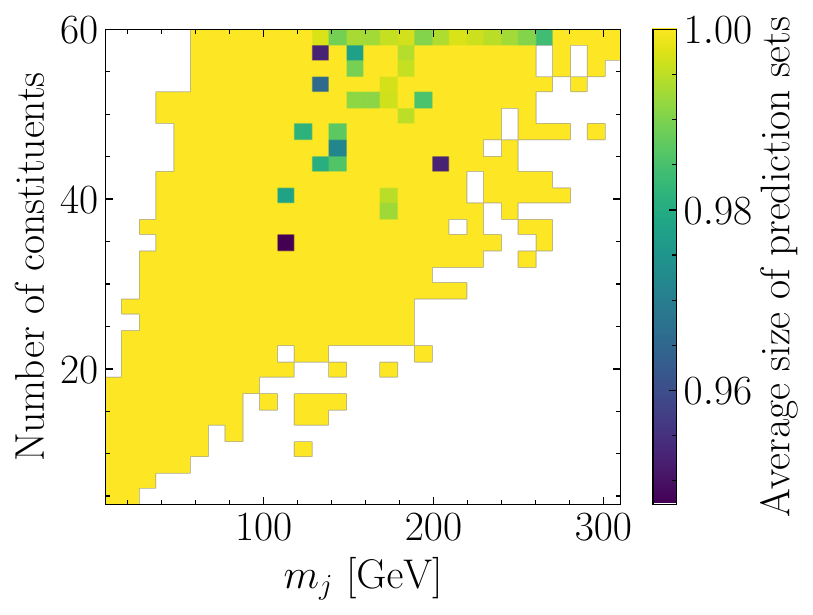}}\qquad
    \subfigure[MB8S\label{fig:mb8s-phasespace}]{\includegraphics[width=.45\linewidth]{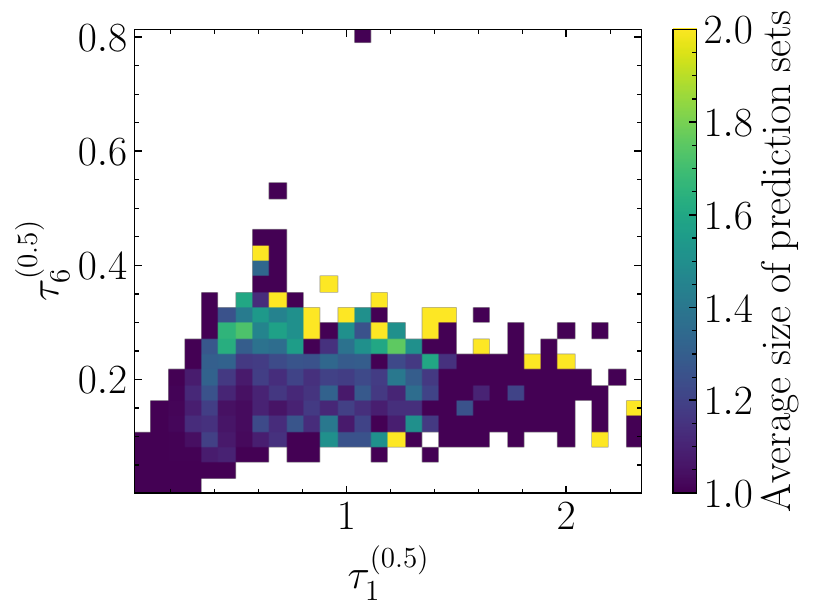}}
    \caption{Binary split–conformal classification at $1-\alpha=0.9$: maps of mean prediction–set size. Colour encodes $\mathbb{E}[\,|\Gamma_\alpha(x)|\,]$ in 2D kinematic bins on the test sample (white: no statistics). \textbf{(a)} PFN and \textbf{(b)} PFIN in the $(m_j,\,N_{\rm const})$ plane; \textbf{(c)} MB8S in the $(\tau_1^{(0.5)},\,\tau_6^{(0.5)})$ plane. For a binary classification $|\Gamma_\alpha(x)|\in\{0,1,2\}$: darker regions indicate confident single–label predictions, brighter regions indicate frequent abstention (both labels included). Prediction sets are built with the score $S(x,y)=1-p_\theta(y\mid x)$ and the quantile index $k=\lceil (n_{\rm cal}+1)(1-\alpha)\rceil$.}
    \label{fig:binary-class-phasespace}
\end{figure}

Figure~\ref{fig:binary-class-phasespace} presents two–dimensional maps of the mean conformal prediction–set size $\mathbb{E}[\,|\Gamma_\alpha(x)|\,]$ at a fixed nominal coverage level of \(1-\alpha=0.9\). For the PFN (Fig.~\ref{fig:pfn-phasespace}) and PFIN (Fig.~\ref{fig:pfin-phasespace}) models, the maps are shown in the physically relevant \((m_j,\,N_{\rm const})\) plane, while for the MB8S baseline (Fig.~\ref{fig:mb8s-phasespace}) a representative pair of high–level N-subjettiness observables \((\tau_1^{(0.5)},\,\tau_6^{(0.5)})\) is used. The colour scale indicates the average prediction–set size per bin, ranging from 0 (unable to label) to 2 (maximal ambiguity for binary classification).

Across all architectures, the prediction–set size exhibits strong dependence on the local structure of phase space. For both PFN and PFIN, regions of large jet mass (\(m_j \gtrsim 200~\mathrm{GeV}\)) combined with high constituent multiplicity are associated with systematically larger prediction sets, indicating reduced classifier confidence. This behaviour is consistent with the increasing overlap of class–conditional distributions in these regions. While both models lose discriminative power in this regime, their failure modes differ: PFN frequently assigns both labels to the prediction set, reflecting intrinsic ambiguity, whereas PFIN more often returns empty or inconclusive predictions, indicating an inability to confidently tag either class. The MB8S baseline, by contrast, shows increased ambiguity primarily at the edges of the high–level observable space, likely driven by limited training statistics and sparsely populated regions rather than by intrinsic model uncertainty.

These phase–space maps highlight a key advantage of conformal prediction: beyond guaranteeing global coverage, the size of the prediction set provides a local, interpretable diagnostic of where and how a classifier becomes uncertain, directly linked to physically meaningful observables.

\begin{figure}[ht]
    \centering
    \includegraphics[width=0.95\linewidth]{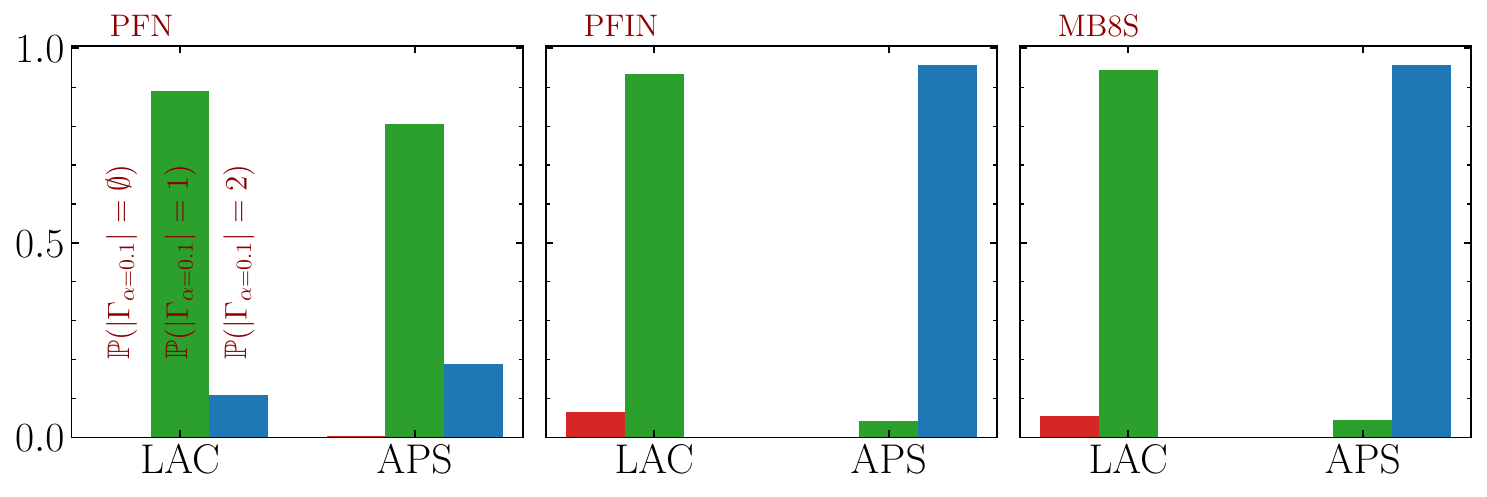}
    \caption{Prediction–set size distribution at $1-\alpha=0.90$ for three binary classifiers (PFN, PFIN, MB8S) under two conformal scores: $1{-}\mathrm{softmax}$ and APS. Bars show the test–set probabilities $P(|\Gamma_\alpha|=\varnothing)$, $P(|\Gamma_\alpha|=1)$, and $P(|\Gamma_\alpha|=2)$. APS yields no empty sets by construction (prefix sets), whereas the $1{-}\mathrm{softmax}$ score can produce a low empty–set rate at this coverage. Differences in the singleton vs two–label rates quantify efficiency: stronger models (PFN, PFIN) produce mostly singletons, while the weaker baseline (MB8S) requires two–label sets more frequently under APS. All results use split CP with quantile index $k=\lceil (n_{\rm cal}+1)(1-\alpha)\rceil$. }
    \label{fig:binary_score-comparisson}
\end{figure}

The choice of nonconformity score is central to the efficiency and interpretability of conformal prediction. Figure~\ref{fig:binary_score-comparisson} compares two widely used score constructions for binary classification: the Least Ambiguous set–valued Classifier (LAC), as defined in Sec.~\ref{sec:class_binary}, and the Adaptive Prediction Set (APS). At a fixed confidence level of 90\%, the red, green, and blue bars represent the probabilities of obtaining prediction sets of size 0, 1, and 2, respectively. The LAC score yields singleton prediction sets predominantly, reflecting a conservative but efficient calibration strategy that preserves sharp predictions whenever possible. In contrast, APS produces a significantly larger fraction of size–2 prediction sets, effectively inflating uncertainty in regions where class probabilities are comparable. While both approaches satisfy the same marginal coverage guarantee and are well-established in the literature, this comparison illustrates that score–function design directly affects the balance between statistical validity and practical usefulness. In particular, overly conservative scores can obscure meaningful structure in the data, underscoring the importance of tailoring the score to the physics problem at hand.

\begin{figure}
    \centering
    \subfigure[PFN, abstentiation rate 11.01\%]{\includegraphics[width=.48\linewidth]{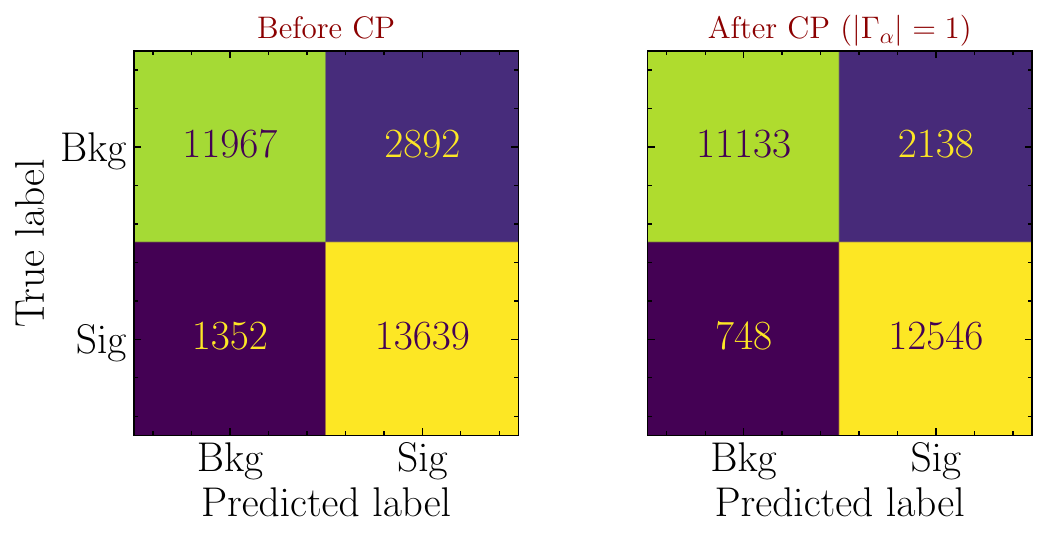}}\quad
    \subfigure[PFIN, abstentiation rate 6.99\%]{\includegraphics[width=.48\linewidth]{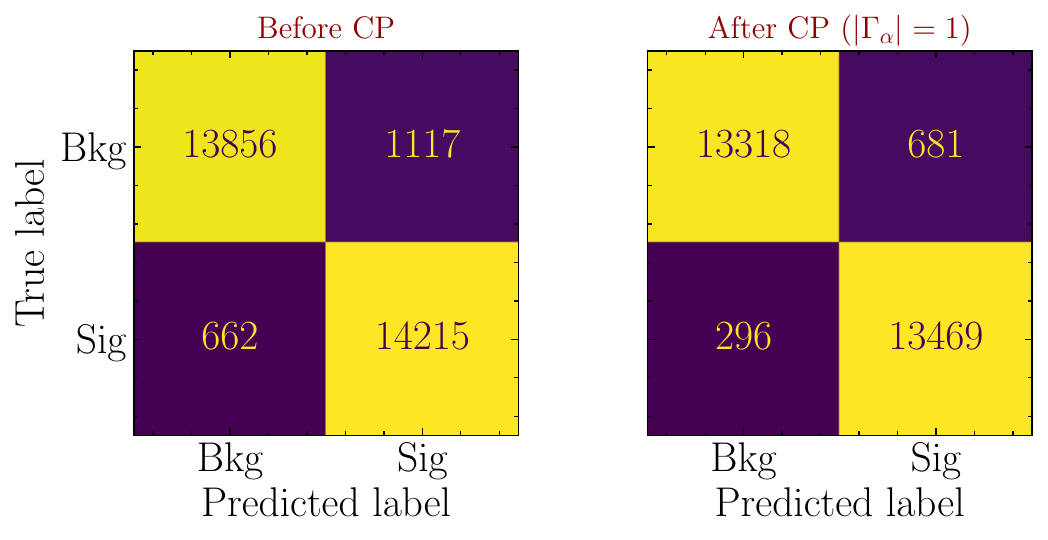}}
    \subfigure[MB8S, abstentiation rate 7.45\%]{\includegraphics[width=.48\linewidth]{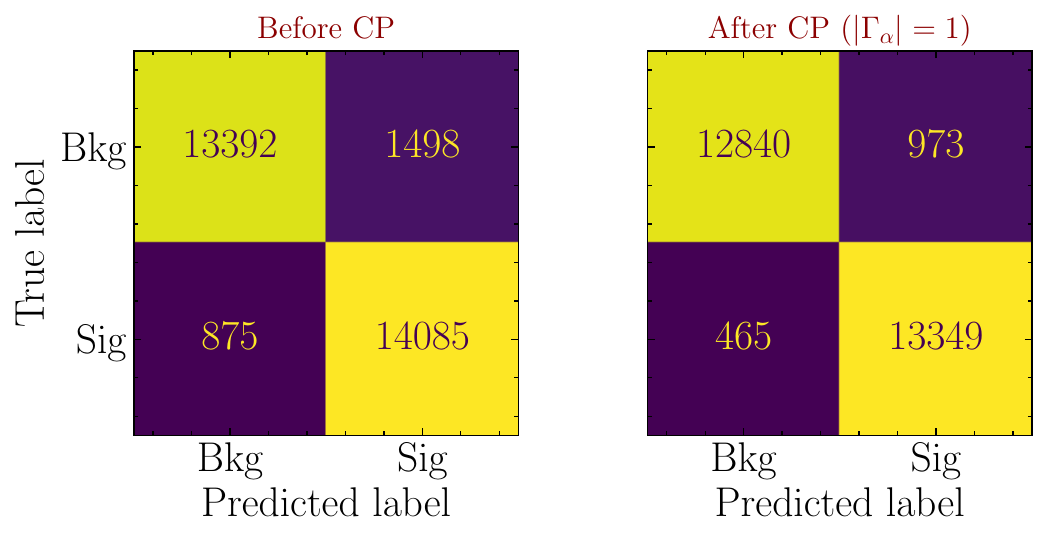}}
    \caption{Confusion matrices before and after split–conformal classification (binary, nominal coverage $1-\alpha=0.90$). Left column: base classifiers (PFN, PFIN, MB8S) evaluated on the full test set. Right column: the same classifiers \emph{after CP}, restricting to events with singleton prediction sets ($|\Gamma_\alpha|=1$); ambiguous events (size $0$ or $2$) are abstained and not counted. Numbers are event counts; the quoted 'abstention rate' is $\mathbb{P}(|\Gamma_\alpha|\neq 1)$ on the test set. After CP, the off–diagonal entries shrink, reflecting fewer confident misclassifications, at the cost of abstaining on a subset of events.}
    \label{fig:confusion-matrix}
\end{figure}

Finally, Fig.~\ref{fig:confusion-matrix} illustrates the impact of conformal calibration on classification performance through confusion matrices shown before and after calibration at a 90\% confidence level. After conformalisation, we restrict attention to events with singleton prediction sets (\(|\Gamma_\alpha|=1\)), corresponding to instances where the classifier is statistically confident in its decision. Across all models, this selection substantially suppresses off–diagonal entries relative to the uncalibrated case, indicating a marked reduction in misclassification rates. This behaviour confirms that conformal prediction acts as an effective uncertainty filter: ambiguous events are systematically withheld from hard classification, while retained predictions exhibit improved reliability. Importantly, this improvement is achieved without retraining the model or altering its internal representations, highlighting conformal prediction as a lightweight yet powerful post–hoc calibration tool.

\section{Anomaly detection}

\subsection{Model specifications}
\label{app:anomaly}

For the application of CP to anomaly detection, we utilised three publicly available models, summarised below.

\begin{itemize}
    \item The \textbf{Deepset-Set VAE} model is a variational autoencoder designed specifically for unordered sets of jet constituents, drawing on the Deep Sets framework to respect inherent permutation invariance. In this architecture, each particle’s feature vector is independently processed by a shared multilayer perceptron (MLP) encoder to produce per-particle embeddings, which are then aggregated via a symmetric pooling operation (e.g., a sum or mean) into a fixed–length global representation. This pooled representation is fed into latent parameter networks that output the mean and variance vectors of a Gaussian latent distribution. During generation, a decoder network reconstructs particle features from samples drawn from the latent space, conditioning on the aggregated set encoding. The Deepset-Set VAE captures set-level statistics and substructure without imposing an arbitrary ordering of constituents. 

    \item The \textbf{Transformer-Clip VAE} extends the VAE framework by incorporating self-attention mechanisms to model interactions between jet constituents before encoding, and by using a clipped objective tailored to efficient anomaly scoring. In this architecture, the input set of particle features is first processed by a transformer encoder that computes context-aware representations for each constituent using multi-head attention layers, enabling the model to capture higher-order correlations across the set. The resulting contextual embeddings are then pooled (e.g., via an attention-weighted sum) to form a global representation that parameterises the VAE’s latent distribution. A corresponding transformer-based decoder reconstructs per-particle features from the latent vectors by attending over learned tokens or latent queries. The "clip" component refers to the way the KL-divergence term in the VAE loss is leveraged as an anomaly score at inference time.

    \item The \textbf{Transformer-set VAE} combines set-structured inputs with a transformer-style encoder–decoder backbone to improve modelling of inter-particle relations while retaining permutation invariance. In this construction, the encoder comprises multiple layers of transformer blocks that apply self-attention across all constituent particle feature vectors, generating contextualised latent embeddings without requiring explicit ordering; permutation invariance is preserved either through shared positional encodings adapted for sets or by operating on unordered inputs with symmetric attention pooling. These encoded features are aggregated into a compact latent representation from which the mean and variance components of a latent distribution are derived. The decoder mirrors this structure, using transformer layers conditioned on latent samples to reconstruct the original set of particle features, reflecting the learned correlations introduced by attention.
\end{itemize}

\subsection{Additional results for conformal prediction in anomaly detection}
\label{app:anomaly-additional}

\begin{figure}[h]
    \centering
    \subfigure[Transformer Clip VAE\label{fig:sig-alpha-transclip}]{\includegraphics[width=0.48\linewidth]{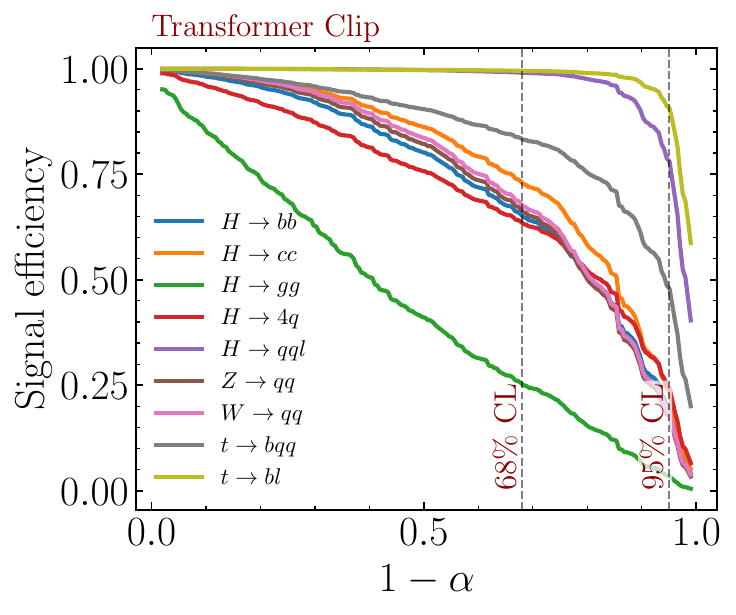}}\quad
    \subfigure[Deepset VAE\label{fig:sig-alpha-deepset}]{\includegraphics[width=0.48\linewidth]{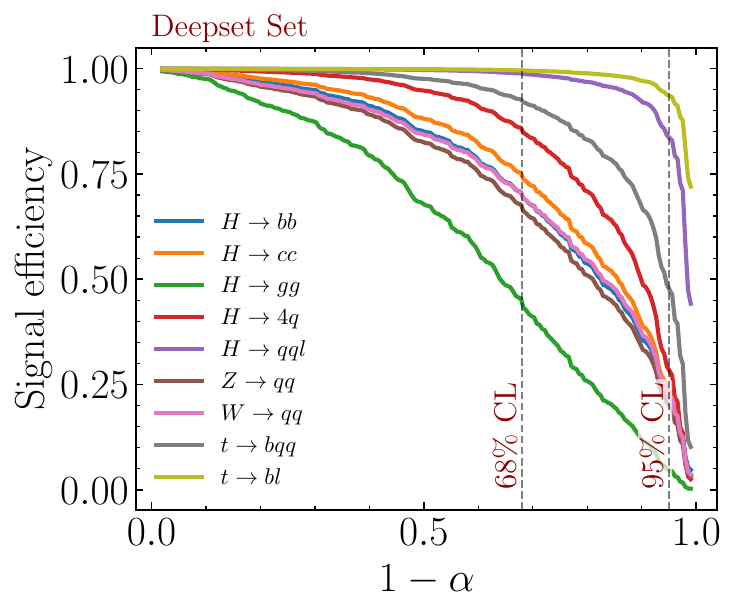}}
    \caption{Signal efficiency as a function of $1-\alpha$ for Transformer Clip and Deepset VAE. Each colour represents efficiency on different signal process.}
    \label{fig:app-sig-alpha}
\end{figure}

In this appendix, we present additional results for the remaining architectures used in the conformal anomaly–detection study, complementing the main results discussed in Sec .~\ref {sec:anomaly}. Figure~\ref{fig:app-sig-alpha} shows the signal efficiency as a function of the nominal coverage level \(1-\alpha\), analogous to Fig.~\ref{fig:anomaly-sig-eff} but now for all anomaly–detection models considered. As in the main text, the calibration threshold is fixed using background-only (QCD) events, such that the false positive rate is controlled at level \(\alpha\).

Across the full range of miscoverage levels, the DeepSets autoencoder exhibits systematically higher signal efficiency than the Transformer-based architectures. Nevertheless, the qualitative dependence of the efficiency on the underlying signal process is remarkably consistent across models. In particular, all architectures achieve higher efficiencies for signals containing leptonic final states, while purely hadronic signals such as \(H\to gg\) remain the most challenging to detect. This behaviour reflects that the leptonic particle content lies farther from the QCD training manifold and is therefore more readily identified as anomalous by all models.

\begin{figure}[h]
    \centering
    \includegraphics[width=.95\linewidth]{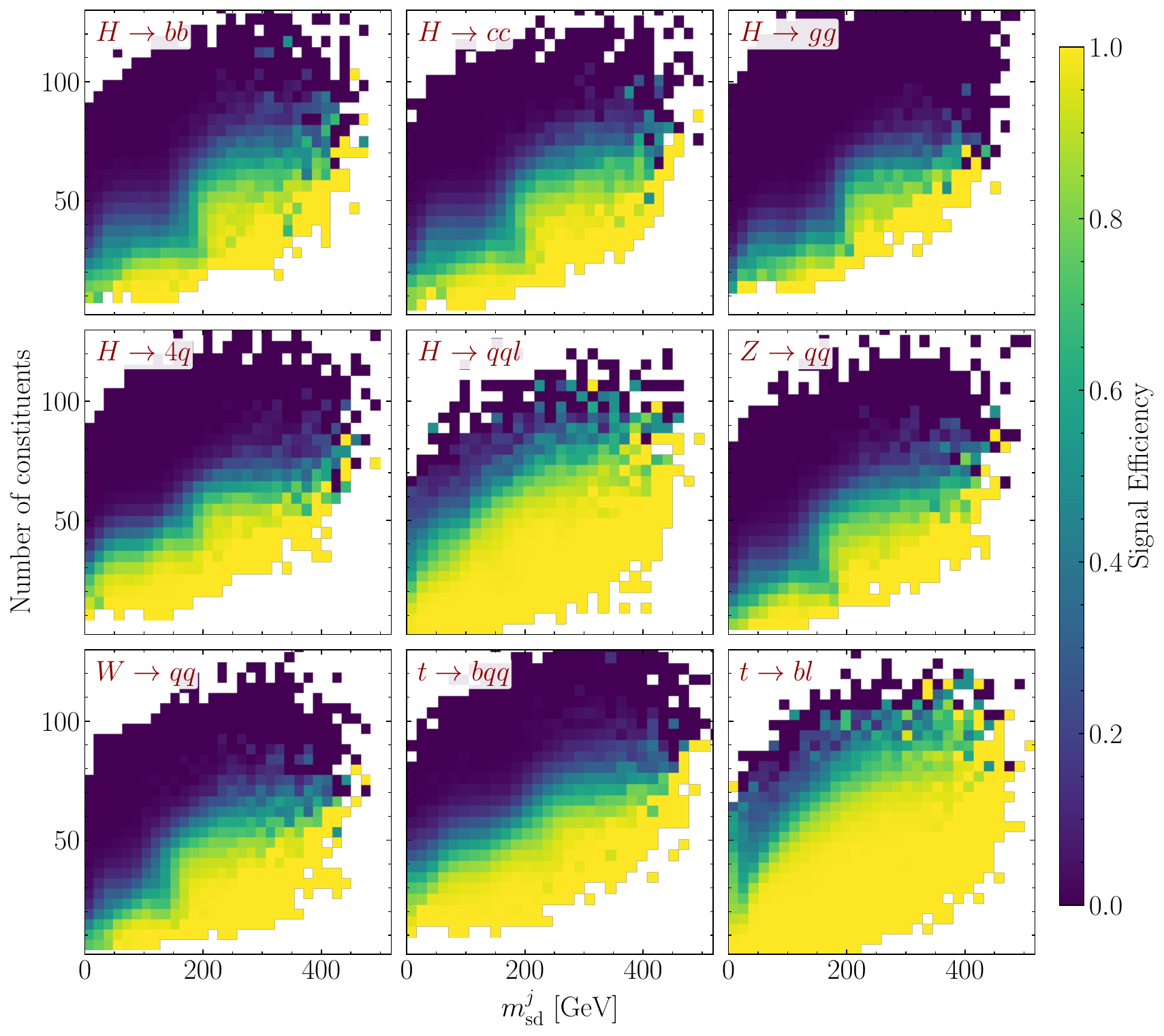}
    \caption{Same figure as Fig.~\ref{fig:kin-anomaly-transformerset} but for Transformer Clip VAE.}
    \label{fig:kin-anomaly-transformerclip}
\end{figure}

\begin{figure}[h]
    \centering
    \includegraphics[width=.95\linewidth]{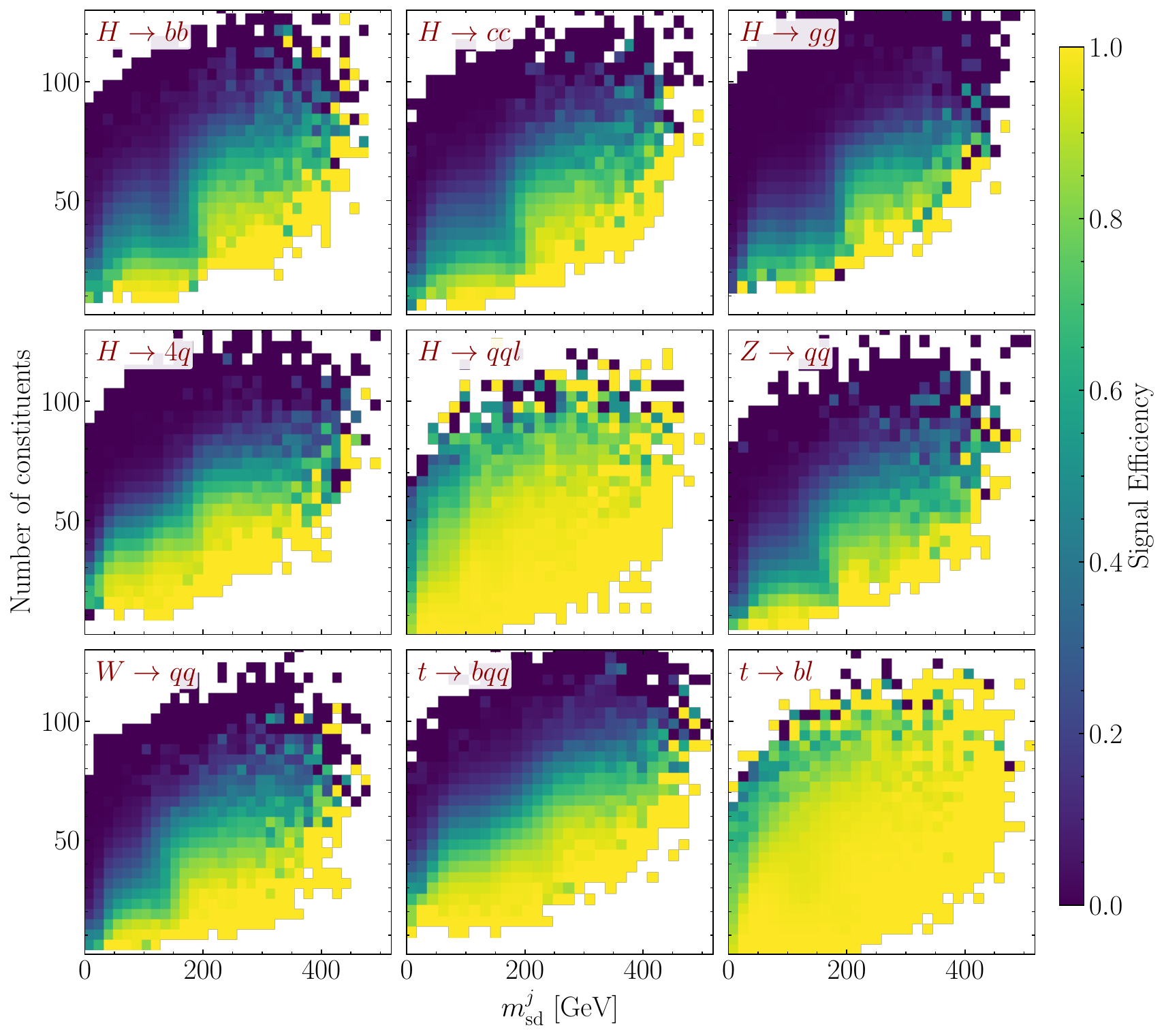}
    \caption{Same figure as Fig.~\ref{fig:kin-anomaly-transformerset} but for Deepset VAE.}
    \label{fig:kin-anomaly-deepset}
\end{figure}

The same trends persist when signal efficiency is examined as a function of kinematic phase space. Figures~\ref{fig:kin-anomaly-deepset} and~\ref{fig:kin-anomaly-transformerclip} show the signal efficiency at 90\% confidence level for the DeepSets and Transformer Clip autoencoders, respectively, mapped onto the plane of soft-drop mass and constituent multiplicity \(N_{\rm const}\). For leptonic signals, the DeepSets model retains high efficiency even in regions of high constituent multiplicity, whereas Transformer-based models tend to lose sensitivity. This suggests that permutation-invariant architectures are more robust to variations in particle multiplicity, whereas attention-based models appear to struggle as the event topology becomes increasingly complex and QCD-like.

From the perspective of conformal prediction, these results highlight two important points. First, the conformal calibration enforces comparable background rejection across architectures, allowing meaningful, like-for-like comparisons of signal efficiency that are not obscured by arbitrary threshold choices. Second, the kinematically resolved efficiency maps reveal where architectural inductive biases most strongly influence anomaly sensitivity, providing actionable guidance for model selection and design in future searches. In this sense, conformal prediction not only provides statistically controlled anomaly decisions but also serves as a powerful diagnostic tool for understanding the strengths and limitations of different models.

\end{document}